\documentclass[12pt]{article}
\usepackage{amssymb,amsmath}
\usepackage{comment}
\usepackage{graphicx}
\usepackage{bm}
\usepackage{color}
\usepackage{here}
\usepackage{enumerate}
\usepackage{subfigure}
\usepackage{hyperref}

\numberwithin{equation}{section}

\begin{document}

\thispagestyle{empty}
\begin{flushright}
DCPT-16/63 \\
NCTS-TH/1611
\\

\end{flushright}
\vskip2cm
\begin{center}
{\Large \bf Mock Modular Index of M2-M5 Brane System}

\vskip1.5cm
Tadashi Okazaki\footnote{tadashiokazaki@phys.ntu.edu.tw} 

\bigskip
{\it 
Department of Physics and Center for Theoretical Sciences,\\
National Taiwan University, Taipei 10617, Taiwan}
\\
\bigskip
and
\\
\bigskip
Douglas J. Smith\footnote{douglas.smith@durham.ac.uk}

\bigskip
{\it Department of Mathematical Sciences, Durham University,\\
Lower Mountjoy, Stockton Road, Durham DH1 3LE, UK}

\end{center}

\vskip1cm
\begin{abstract}
We present BPS indices of the supergroup WZW models 
that live on intersecting M2-M5 brane systems. 
They can encode data of the stretched M2-branes between M5-branes 
and count the BPS states. 
They are generally expressed in terms of mock theta functions  
via the Kac-Wakimoto character formula of the affine Lie superalgebra. 
We give an explicit expression of the index for the $PSL(2|2)_{k=1}$ WZW model 
in terms of the second order multi-variable Appell-Lerch sum. 
It indicates that wall-crossing occurs in the BPS state counting due to the $C$-field on the M5-branes. 
\end{abstract}

\newpage
\setcounter{tocdepth}{2}
\tableofcontents

\section{Introduction}
\label{secintro}
We recently proposed that a particular topologically twisted field theory 
arising from the intersection of M2-branes and M5-branes 
is described by a supergroup WZW model \cite{Okazaki:2015fiq}. 
In this paper we explore this description and in
particular we present a BPS index for such a WZW model. 
It encodes data of the stretched M2-branes between the M5-branes. 
When all the M2-branes are suspended between the M5-branes, the index can be computed via
the Kac-Wakimoto character formula \cite{MR1810948, MR3200431}. 
This gives an explicit expression for the index 
in the case of supergroup $PSL(2\vert 2)$. 
The result can be expressed in terms of Jacobi theta functions 
and second order multi-variable Appell-Lerch sums. 
However, while it is holomorphic it is not modular. 
Based on Zwegers' method \cite{Zwegers:2008zna} 
and results of Dabholkar et al.\ \cite{Dabholkar:2012nd} 
and Ashok et al.\ \cite{Ashok:2014nua}
we demonstrate how to complete the expressions to give a modular index, 
which would contribute to the torus partition function. 
The Appell-Lerch sum of order 2 in the counting function is suggestive of 
the occurrence of wall-crossing phenomenon due to the dependence of the
Fourier expansion on the parameter region.
The structure of the paper is as follows.

In Section~\ref{secm2m5} we review
some background material, including notation for supergroups and summarize our
previous work \cite{Okazaki:2015fiq}. 
Then we review the main result of that paper, that after topological twisting, a certain configuration of M2-branes
stretched between M5-branes gives rise to a supergroup WZW model. 
We also comment on type IIB brane configurations related to these M-branes configurations.

In Section~\ref{secaff} we review properties of affine Lie superalgebras which are relevant to the supergroup WZW index derived in this paper. As well as
defining notation, we discuss the important concept of atypical modules. 
In Section~\ref{secdiagram} the connection between these atypical modules and
brane configurations is explained. This relation for M-branes is similar to the
relation proposed by Mikhaylov and Witten \cite{Mikhaylov:2014aoa} for branes
in type IIB. We discuss the connection between these realizations of atypical
modules from brane configurations.

Section~\ref{secmock} contains the main result of this paper, the derivation of
an index for the supergroup WZW models. The details of the index are explained,
including an explicit evaluation for the case of supergroup $PSL(2|2)$. The
result is a holomorphic but not modular expression.
In Section~\ref{secALsums} we adapt results in \cite{Ashok:2014nua} to find
the modular completion of this index. We comment on the relation to wall-crossing 
in counting of the BPS states of the M2-M5 system and black hole microstates. 
In Section~\ref{secdis} we summarize our results and discuss future directions.

\section{M2-M5 System and Supergroup WZW Model}
\label{secm2m5}
We start with some preliminaries, reviewing some essential properties of
supergroups before summarizing our previous results. In particular we briefly
review the M2-M5 branes construction and the resulting supergroup WZW
model. We also give a description of type IIB brane configurations related to
these M-brane configurations through compactification and T-duality.
\subsection{Preliminaries}
\label{liesuper}
To formulate our result in detail, we first fix our notation and conventions. 
Let $\mathfrak{sg}=\mathfrak{g}_{\overline{0}}\oplus \mathfrak{g}_{\overline{1}}$ be the Lie superalgebra 
where $\mathfrak{g}_{\overline{0}}$ and $\mathfrak{g}_{\overline{1}}$ are respectively 
the even and odd parts of the superalgebra $\mathfrak{sg}$. 
The bilinear form $( \cdot,\cdot): \mathfrak{sg}\otimes \mathfrak{sg}\rightarrow \mathbb{C}$ 
obeys the following properties \cite{MR0486011}
\begin{align}
\label{bi1}
( a,b)&=0\ \ \ \ \ \textrm{for $a\in \mathfrak{g}_{\overline{0}}, b\in \mathfrak{g}_{\overline{1}}$}, 
& (\textrm{even})\\
\label{bi2}
(a,b)&=(-1)^{\deg a\cdot \deg b}( b,a ), 
& (\textrm{supersymmetric})\\
\label{bi3}
( [a,b],c )&=
( a,[b,c] )
& (\textrm{invariant})
\end{align}
and the Lie superbracket 
$[\cdot, \cdot]: \mathfrak{sg}\otimes \mathfrak{sg}\rightarrow \mathfrak{sg}$ 
satisfies the following axioms \cite{MR0486011}
\begin{align}
\label{bra1}
[a,b]&=ab+(-1)^{\deg a\cdot \deg b}ba,\\
\label{jacobi1}
[a,[b,c]]&=[[a,b],c]
+(-1)^{\deg a\cdot \deg b}[b,[a,c]]
\end{align}
where we have assigned the grade such that $\deg a=0$ for $a\in \mathfrak{g}_{\overline{0}}$ 
while $\deg a=1$ for $a\in \mathfrak{g}_{\overline{1}}$. 
The relation (\ref{jacobi1}) is the Jacobi identity.

Let $\mathfrak{h}$ be the Cartan subalgebra of $\mathfrak{sg}=\mathfrak{gl}(N|M)$ 
which is a set of diagonal matrices with basis
$ 
\left\{
E_{1,1},\cdots, E_{N,N}; E_{N+1,N+1},\cdots, E_{N+M,N+M}
\right\}
$ 
where $E_{ij}$ is the matrix whose entries are all zero except for the
$ij$-entry which is one.

Let $\{\epsilon_{1},\cdots, \epsilon_{N}; \delta_{1},\cdots, \delta_{M}\}$ be the basis for the dual space $\mathfrak{h}^{*}$. 
The bilinear form on $\mathfrak{h}^{*}$ can be defined by 
$(\epsilon_{i},\epsilon_{j})=-(\delta_{i},\delta_{j})=\delta_{ij}$ and $(\epsilon_{i},\delta_{j})=0$. 
We denote a set of roots by 
$\Delta=\Delta_{\overline{0}}\cup \Delta_{\overline{1}}$, 
where 
\begin{align}
\label{root1}
\Delta_{\overline{0}}&=
\left\{
\epsilon_{i}-\epsilon_{j}|1\le i\neq j\le N
\right\} 
\cup 
\left\{
\delta_{k}-\delta_{l}|1\le k\neq l \le M
\right\}, \\
\Delta_{\overline{1}}&=
\left\{
\epsilon_{i}-\delta_{k}|1\le i \le N, 1\le k \le M
\right\}. 
\end{align}

Simple roots are the elements $\alpha_{i}\in \mathfrak{h}^{*}$ 
that obey  $\alpha_{i}(h_{j})=a_{ij}$ where $A=(a_{ij})$ is a symmetrized Cartan matrix. 
Let $\Pi$ be a set of simple roots, 
$Q:=\mathbb{Z}\Pi$ be the root lattice and $Q^{+}=\mathbb{Z}_{\ge 0}\Pi$. 
We define a set of positive roots by $\Delta^{+}=\Delta \cap Q^{+}$. 
%
%
%
%
%
%
A set $\Pi$ of simple roots specifies the decomposition 
of $\Delta$ into positive and negative roots $\Delta=\Delta^{+}\cup \Delta^{-}$ 
and the Borel decomposition 
$\mathfrak{sg}=\mathfrak{n}^{+}\oplus \mathfrak{h}\oplus \mathfrak{n}^{-}$ 
where $\mathfrak{b}=\mathfrak{h}\oplus \mathfrak{n}^{+}$ is the Borel subalgebra 
and $\mathfrak{n}^{\pm}=\bigoplus_{\alpha\in \Delta^{+}}
\mathfrak{sg}_{\pm\alpha}$ 
with $[\mathfrak{h},\mathfrak{n}^{+}]\subset \mathfrak{n}^{+}$, 
$[\mathfrak{h},\mathfrak{n}^{-}]\subset \mathfrak{n}^{-}$.

The Weyl vector is defined by 
$\rho=\frac12 \sum_{\alpha \in \Delta_{\overline{0}}^{+}}\alpha -\frac12\sum_{\alpha\in \Delta_{\overline{1}}^{+}}\alpha$ 
and it depends on a choice of the set of positive roots. 
%
%
%
%
%
%
The Weyl group $W\subset GL(\mathfrak{h}^{*})$ of $\mathfrak{sg}$ 
is the group generated by the reflections 
$r_{\alpha}(\Lambda)=\Lambda-\frac{2(\alpha,\Lambda)\alpha}{(\alpha,\alpha)}$ 
with respect to non-isotropic roots $\alpha\in \Delta_{\overline{0}}$. 
Let $h^{\vee}$ be the dual Coxeter number, 
i.e.\ half of the eigenvalue of the Casimir operator associated to the bilinear form $(\cdot, \cdot)$.
For $h^{\vee}\neq 0$ we define \cite{MR1810948}
\begin{align}
\label{rootsharp}
\Delta_{\overline{0}}^{\sharp}&:=
\left\{
\alpha\in \Delta_{\overline{0}}|h^{\vee}(\alpha,\alpha)>0
\right\},& 
W^{\sharp}&:=
\left\{
r_{\alpha}\in W|\alpha\in \Delta_{\overline{0}}^{\sharp}
\right\}.
\end{align}
For $h^{\vee}=0$, i.e. $\mathfrak{sg}=\mathfrak{gl}(N|N)$, $\mathfrak{osp}(2N+2|2N)$ and $D(2,1;\alpha)$,  
the root system $\Delta_{\overline{0}}$ is a union of two orthogonal root subsystem. 
For $\mathfrak{gl}(N|N)$ we define
$\Delta_{\overline{0}}^{\sharp}=\mathfrak{gl}(N)$ 
\cite{MR1327543, MR2866851}. 
%
%
%
%
%
%
%
We set
\begin{align}
\label{sroot2a}
\overline{\Delta}_{\overline{0}}
&=\{\alpha\in \Delta_{\overline{0}}\ |\ \frac12 \alpha\notin \Delta\}, &\overline{\Delta}_{\overline{1}}
&=\{\alpha\in \Delta_{\overline{1}}\ |\ (\alpha,\alpha)=0 \}
\end{align}
and define 
\begin{align}
\label{sign1}
\mathrm{sgn}^{+}(w)
&:=(-1)^{l(w)},& 
\mathrm{sgn}^{-}(w)
&:=(-1)^{m(w)}
\end{align}
where $l(w)$ is the length function on $W$, 
i.e. the number of reflections with respect to roots from 
$\Delta_{\overline{0}}^{+}$ appearing in the decompositions of $w\in W$,
and $m(w)$ is the number of reflections for the realization of $w$ from $\overline{\Delta}_{\overline{0}}^{+}$. 
In terms of the isotropic root $\beta\in \Pi$, 
we can define an odd reflection by \cite{MR1201236}
\begin{align}
\label{oddr1}
r_{\beta}(\Delta^{+})
&=(\Delta^{+} \setminus \{\beta\})\cup \{-\beta\} 
\end{align}
and it is also a set of simple roots for $\mathfrak{sg}$ \cite{MR2743764}. 
It turns out that 
any two sets of positive roots can be obtained 
from each other by applying a finite sequence of odd reflections.

A weight $\lambda\in \mathfrak{h}^{*}$ is called dominant 
if $\frac{2(\lambda,\alpha)}{(\alpha,\alpha)}\ge 0$ for all $\alpha\in \Delta_{\overline{0}}^{+}$, 
strictly dominant if $\frac{2(\lambda,\alpha)}{(\alpha,\alpha)}>0$ for all $\alpha\in \Delta_{\overline{0}}^{+}$ 
and integral if $\frac{2(\lambda,\alpha)}{(\alpha,\alpha)}\in \mathbb{Z}$ for all $\alpha\in \Delta_{\overline{0}}^{+}$. 
Let $P$ be a set of integral weights and 
$P^{+}$ be a set of dominant integral weight. 
We define 
\begin{align}
\label{diweight1}
\mathbb{P}^{+}&=
\left\{
\lambda\in P^{+}|
(\lambda+\rho,\epsilon_{i})\in\mathbb{Z}, 
(\lambda+\rho,\delta_{k})\in \mathbb{Z}
\right\}.
\end{align}
$\mathbb{P}^{+}$ does not depend on a choice of $\Pi$.

\subsection{M2-M5 system}
\label{subsecm2m5}
The starting point for the brane construction is a set of $N$ M2-branes
suspended between two M5-branes. As is well known the description of multiple
M2-branes is given by supersymmetric Chern-Simons matter theories, 
either the BLG or ABJM model.
In fact the configuration of a fuzzy funnel of M2-branes producing an M5-brane
is described by the Basu-Harvey equation \cite{Basu:2004ed},
a generalization of the Nahm equation. Requiring this to be a BPS equation
of the M2-brane theory was a crucial ingredient used by Bagger and Lambert
\cite{Bagger:2006sk} in the derivation of the
supersymmetry transformations, leading to the
BLG model.
Such BPS equations were already studied in the context of the a variety of M2-M5 systems
\cite{Nogradi:2005yk, Berman:2006eu}, including a generalization of the
Basu-Harvey equation by Berman and Copland \cite{Berman:2005re}
which was shown to apply to
the BLG model in \cite{Krishnan:2008zm}. 
However, as discussed in \cite{Nastase:2009ny}
it is not clear how the required geometry, a funnel with fuzzy 3-sphere
cross-section, can be realized for arbitrary numbers of M2-branes, although
BPS configurations of the BLG or ABJM models describing the M2-M5 or D2-D4 systems have also been
discussed in detail in \cite{Hanaki:2008cu, Nosaka:2012tq}.

When describing open M2-branes by the BLG or ABJM action, a crucial feature is that
when we have a boundary, a Chern-Simons term will give rise to a
WZW model. On the other hand, the boundary of M2-branes on M5-branes corresponds
to the self-dual strings in the M5-brane theory. 
The description of such
systems has been considered in terms of boundary conditions for the Chern-Simons theories, and through adding boundary degrees of freedom to restore the
gauge symmetry of the Chern-Simons theory in the presence of a boundary
\cite{Berman:2009kj, Chu:2009ms, Berman:2009xd, Faizal:2011cd, Hosomichi:2014rqa, Armoni:2015jsa, Niarchos:2015lla, Faizal:2016skd}.

In \cite{Okazaki:2015fiq}, with the aim of describing 
the internal dynamics of these strings, we choose a brane configuration in order
to project out the transverse scalar degrees of freedom, and to decouple the
two-dimensional boundary theory from the `bulk' three-dimensional M2-brane
world-volume theory. Another motivation was the construction of
Mikhaylov and Witten \cite{Mikhaylov:2014aoa} building on results in
\cite{Gaiotto:2008sa, Gaiotto:2008sd, Kapustin:2009cd}
studying field theories in one higher dimension. 
There, four-dimensional twisted $\mathcal{N}=4$ SYM theory 
with a boundary was shown to give rise to a three-dimensional Chern-Simons
theory with a supergroup. 

\begin{figure}
\begin{center}
\includegraphics[width=10cm]{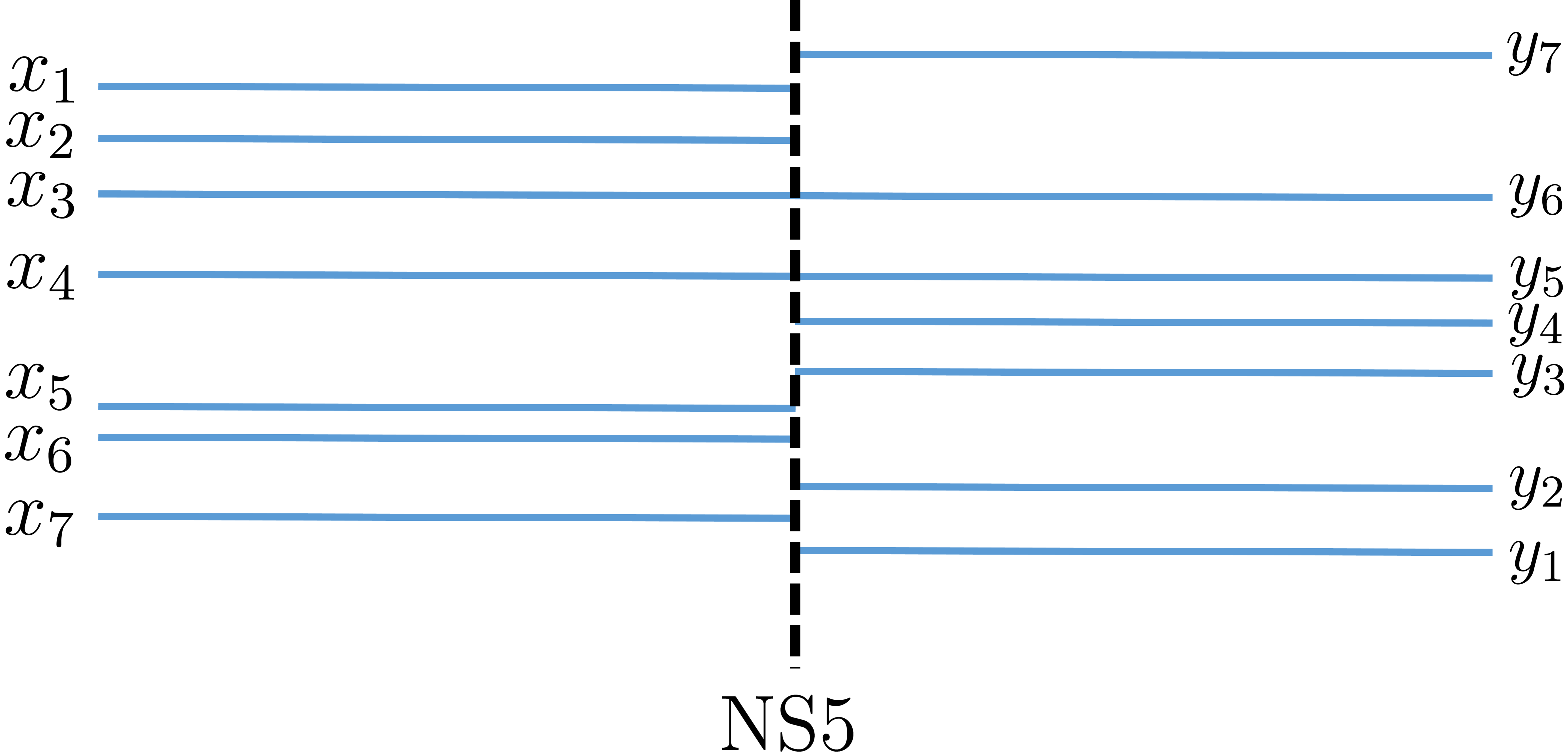}
\caption{A brane configuration with $N$ D3-branes and $M$ D3-branes terminating on 
a single NS5-brane at $x^{6}=0$ from left and right respectively.
The horizontal lines are the D3-branes extending in the $x^{6}$ direction.  
The sequences $\{x_{i}\}$ and $\{y_{k}\}$ of the heights of the D3-branes in the picture 
label the RR charges of the D3-branes. 
In this example $N=M=7$. }
\label{mw2}
\end{center}
\end{figure}

In the type IIB setting this is realized for D3-branes ending on both sides of
a single NS5-brane (see Figure \ref{mw2}).
When $N$ D3-branes end on one side $(x^{6}<0)$ of a single NS5-brane at $x^{6}=0$, 
and $M$ D3-branes on the other side $(x^{6}>0)$, 
the system supports 
four-dimensional $\mathcal{N}=4$ $U(N)$ SYM theory for $x^{6}<0$ 
and $U(M)$ SYM theory for $x^{6}>0$. 
With an appropriate choice of supercharges $\mathcal{Q}$ via topological twist, 
the complete action of the effective theory is shown to be written as a sum of 
a $\mathcal{Q}$-exact term and a $U(N|M)$ supergroup Chern-Simons theory at 
the common boundary at $x^{6}=0$
\begin{align}
\label{scs1}
S&=\left\{\mathcal{Q}, \cdots\right\}
+\frac{i\mathcal{K}}{8\pi}
\int \mathrm{Str}
\left(
\mathcal{A}\wedge d\mathcal{A}
+\frac23 \mathcal{A}\wedge \mathcal{A}\wedge \mathcal{A}
\right)
\end{align}
where $\mathcal{A}$ is a $\mathfrak{u}(N|M)$-valued field and 
$\mathcal{K}$ is a complex parameter.

In order to meet the counterpart of the above construction in M-theory we considered the M2-branes to be
suspended between two differently oriented M5-branes, labelled M5 and M5$'$,
which share a four-dimensional world-volume. The details are summarized as:
\begin{align}
\label{M2M5M5p}
\begin{array}{cccccccccccc}
&0&1&2&3&4&5&6&7&8&9&10\\
\textrm{M5}
&\circ&\circ& &\circ&\circ& & & & &\circ&\circ\\
\textrm{M5$'$}
&\circ&\circ& & & &\circ&\circ& & &\circ&\circ\\
\textrm{M2}
&\circ&\circ&\circ& & & & & & & & 
\end{array}
\end{align}
where $\circ$ are the directions spanned by the branes.

Taking the limit where the
separation between the M5- and M5$'$-branes vanishes would produce a two-dimensional
theory but still with two transverse scalar degrees of freedom corresponding to
the freedom of the M2-branes to move in the $(x^{9},x^{10})$ directions. In order to
relate the standard Chern-Simons matter theory to a supergroup WZW model we first
implement a topological twist. 
We consider the theory in the Euclidean space with the
$(x^{0}, x^{1}, x^{9}, x^{10})$ directions a K3 manifold, 
and the M2-brane wrapping a Riemann surface
$\Sigma \subset K3$. Then we twist the theory by identifying the
twisted two-dimensional Euclidean Lorentz group as 
$$ SO(2)_E' = \textrm{diag}(SO(2)_E \times SO(2)_R) $$
where $SO(2)_E$ is the Euclidean Lorentz group on the two-dimensional Riemann
surface and $SO(2)_R$ is the rotation group in the $(x^{9}, x^{10})$ directions  (see \cite{Okazaki:2014sga} for details).

Now it turns out that in the twisted theory the fields combine, with the result
that the Chern-Simons matter theory becomes a Chern-Simons theory with
complexified gauge fields. The boundary action then becomes a WZW model with
bosonic part described by the complexified gauge group, i.e.\
$SL(2, \mathbb{C}) \times SL(2, \mathbb{C})$ for the BLG theory. However, the
fermionic fields couple in such a way that the complete description is in terms
of a supergroup. In other words in this construction the two groups on the
boundary are identified together as the even part of a supergroup. 
Specifically, for the BLG theory we arrive at a boundary $PSL(2\vert 2)$ WZW model. This
theory is summarized in the following section. Of course, this can also be
viewed as a special case arising from the ABJM action.
However, note that in
detail while $SU(2) \times SU(2) \rightarrow SL(2, \mathbb{C}) \times SL(2, \mathbb{C}) \rightarrow PSL(2 \vert 2)$, with gauge group $U(N) \times U(N)$ we have
$U(N) \times U(N) \rightarrow GL(N, \mathbb{C}) \times GL(N, \mathbb{C}) \rightarrow GL(N \vert N)$. 

\begin{figure}
\begin{center}
\includegraphics[width=7cm]{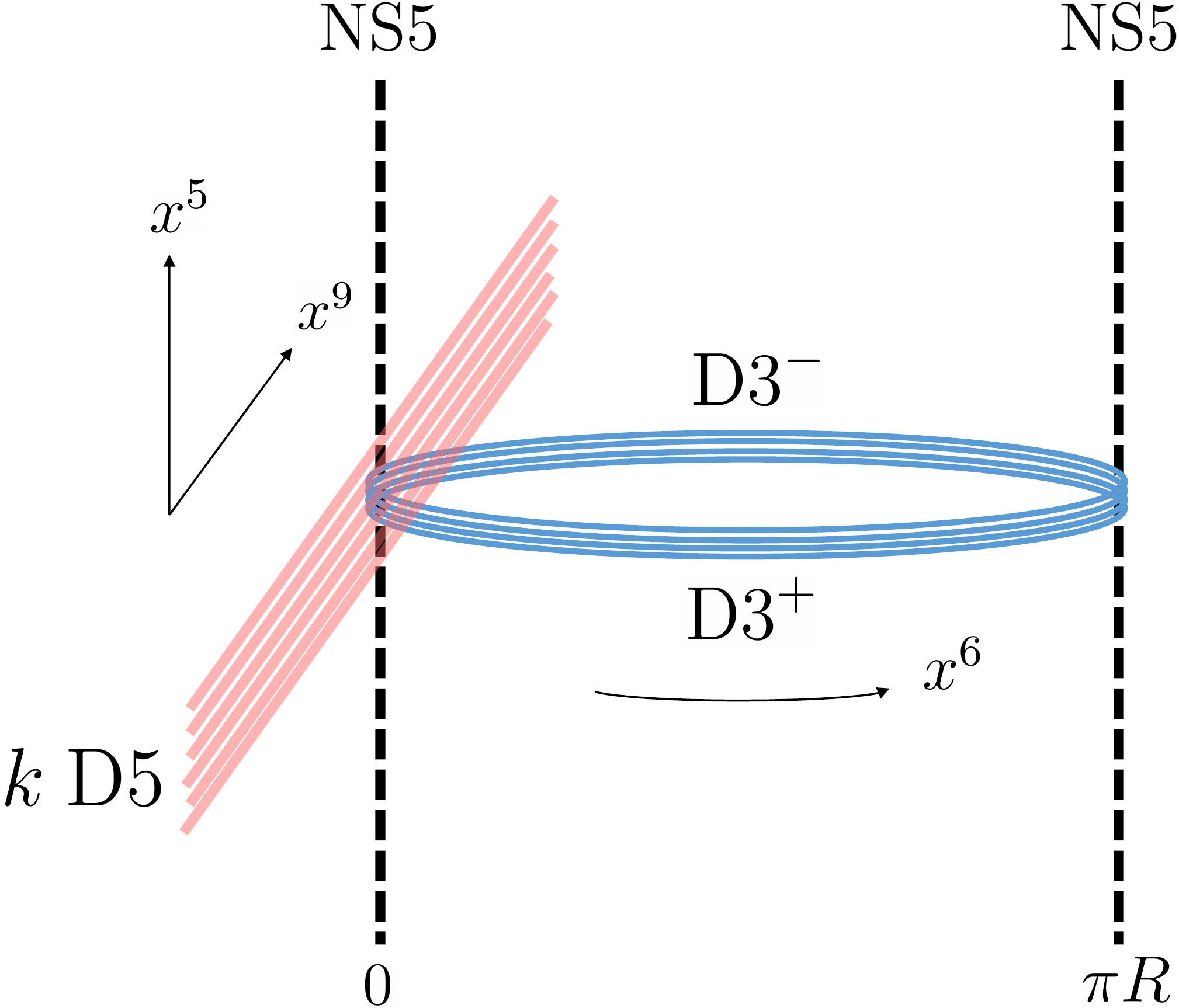}
\caption{The type IIB configuration of the ABJM model. 
Here the $x^{6}$ direction is compact 
and there are two stacks of D3-branes on the circle labelled by D3$^{+}$ and D3$^{-}$. 
}
\label{figabjm1}
\end{center}
\end{figure}

Before proceeding, we also note that the ABJM theory can be seen to arise from
a type IIB brane configuration. The basic connection is that D3-branes wrapped
on a circle T-dualize to D2-branes in type IIA and then lift to M2-branes.
However, to get a Chern-Simons theory rather than SYM theory, 
the D3-branes are taken to intersect two NS5-branes 
at points on this circle, and furthermore $k$
D5-branes also intersect at the position of one of the NS5-branes. This is summarized as
\begin{align}
\label{D3NS5D5}
\begin{array}{ccccccccccc}
&0&1&2&3&4&5&6&7&8&9\\
\textrm{NS5}
&\circ&\circ&\circ&\circ&\circ&\circ & & & & \\
\textrm{D5}
&\circ&\circ&\circ&\circ&\circ& & & & &\circ \\
\textrm{D3}^{+}
&\circ&\circ&\circ& & & &+ && & \\
\textrm{D3}^{-}
&\circ&\circ&\circ& & & &- && & 
\end{array}
\end{align}
The $x^{6}$ direction is taken to have period $2\pi R$ and the two NS5-branes
are located at $x^{6}= 0$ and $x^{6}= \pi R$. The D3-branes split into two stacks
of D3-branes suspended between the NS5-branes, each stack covering one half of
the circle and distinguished by $\pm$ in the above table. The D5-branes are
located at $x^{6}=0$. Note that here the D3-branes are free to move in the $(x^{3},x^{4})$
directions (see Figure \ref{figabjm1}). 

Finally, the intersection of the NS5-brane with the $k$ D5-branes is deformed
to form a $(p,q)$-5-brane web in the $(x^{5}, x^{9})$-plane. Specifically, the
parts of the NS5- and D5-branes with positive $x^{5}$ and $x^9$ are separated
from the parts
with negative $x^{5}$ and $x^{9}$. The two `corners' are then linked by a
$(1,k)$-5-brane with a suitable orientation in the $(x^{5},x^{9})$-plane to preserve
supersymmetry.
As explained in \cite{Aharony:2008ug, Bergman:1999na, Kao:1995gf}
this gives rise to a SYM theory with massive fundamental chiral multiplets,
and integrating those out produces a Chern-Simons theory.

T-dualizing the brane configuration along $x^{6}$ and then lifting to eleven
dimensions results in M2-branes 
with world-volume directions $(x^{0},x^{1},x^{2})$ while the
5-branes become KK-monopoles and D6-branes in type IIA, both of which arise
from KK-monopoles in eleven dimensions. The resulting low energy background is
given by a $\mathbb{Z}_k$ orbifold in the $\mathbb{C}^4$ transverse to the M2-branes.

It is interesting to observe that the type IIB brane origin of ABJM theory 
contains D3-branes ending on an NS5-brane. As shown by Mikhaylov and Witten 
\cite{Mikhaylov:2014aoa}, after topological twisting this intersection gives rise to a 
supergroup Chern-Simons theory at the intersection of the D3-branes with the 
NS5-brane. In the case, as here, with $N$ D3-branes on either side of the 
NS5-brane this can be interpreted as a codimension one defect in the four-dimensional 
$\mathcal{N} = 4$ SYM theory, and at the defect we have a $U(N|N)$ 
supergroup Chern-Simons theory. It is tempting to speculate that the 
appearance of a supergroup in this way is related to the supergroup WZW 
model arising in \cite{Okazaki:2015fiq}. However, the precise link is not clear 
as in the case of M2-branes ending on an M5-brane, the supergroup theory 
arose due to the boundary for the M2-branes. In particular the result did not 
require a supergroup Chern-Simons theory. However, it is certainly the case that 
the structure of the ABJM model is constrained, e.g.\ the conditions for 
such Chern-Simons matter theories to preserve large amounts of supersymmetry 
can be expressed in terms of supergroups \cite{Hosomichi:2008jb, Gaiotto:2008sd, deMedeiros:2008zh}. 

Now in order to relate to an M-theory configuration with M5-branes we need to 
introduce additional 5-branes in the type IIB configuration. 
This has been discussed in the similar context of M2-branes between parallel M5-branes 
by Niarchos \cite{Niarchos:2015lla}. 
Of course, in the case of parallel 5-branes the BPS index for the M-strings has been calculated in \cite{Haghighat:2013gba} 
using various techniques including topological strings. 
However, the type IIB construction as discussed by Niarchos can be used to provide an explicit 
Lagrangian description, albeit without all supersymmetry manifest. 

In our case the following additional D5-branes will give rise to the M5 and
M5$'$ branes in eleven dimensions:
\begin{align}
\label{D5D5p}
\begin{array}{ccccccccccc}
&0&1&2&3&4&5&6&7&8&9\\
\textrm{D5}
&\circ&\circ& &\circ&\circ&\circ  &\circ& & & \\
\textrm{D5$'$}
&\circ&\circ& & & &\circ&\circ&\circ&\circ &
\end{array}
\end{align}
The complete brane configuration in type IIB now preserves two supercharges.
However, this is not quite the right configuration as in the
eleven-dimensional configuration there is an obvious discrete symmetry relating the
M5- and M5$'$-branes. In type IIB we see that the D5-brane shares 
the world-volume directions
$x^3$ and $x^4$ with the NS5- and $(1,k)$-5-branes while the D5{$'$}-brane has a lower
dimensional set of common directions. However, we can maintain this symmetry
in the type IIB configuration by taking the D5-brane to have
embedding $w_1 = w_2$ while the D5$'$-brane has $w_1 = -w_2$, where we define
$w_1 = x^3 + ix^4$ and $w_2 =x^{7}+ ix^{8}$. This preserves exactly the same
supersymmetries in type IIB, while in eleven dimensions this just corresponds
to a change of coordinates. We can therefore schematically describe the D5- and
D5$'$-branes embeddings as
\begin{align}
\label{D5D5p_2}
\begin{array}{ccccccccccc}
&0&1&2&3&4&5&6&7&8&9\\
\textrm{D5}
&\circ&\circ& & \nearrow & \nearrow &\circ&\circ &\nearrow &\nearrow  & \\
\textrm{D5$'$}
&\circ&\circ& & \searrow & \searrow &\circ&\circ &\searrow &\searrow &
\end{array}
\end{align}
and these D5-branes would correspond to the following M5-branes
\begin{align}
\label{DM5M5p_2}
\begin{array}{cccccccccccc}
&0&1&2&3&4&5&6&7&8&9&10\\
\textrm{M5}
&\circ&\circ& & \nearrow & \nearrow &\circ& &\nearrow &\nearrow & &\circ \\
\textrm{M5$'$}
&\circ&\circ& & \searrow & \searrow &\circ& &\searrow &\searrow & & \circ
\end{array}
\end{align}

\begin{figure}
\begin{center}
\includegraphics[width=15cm]{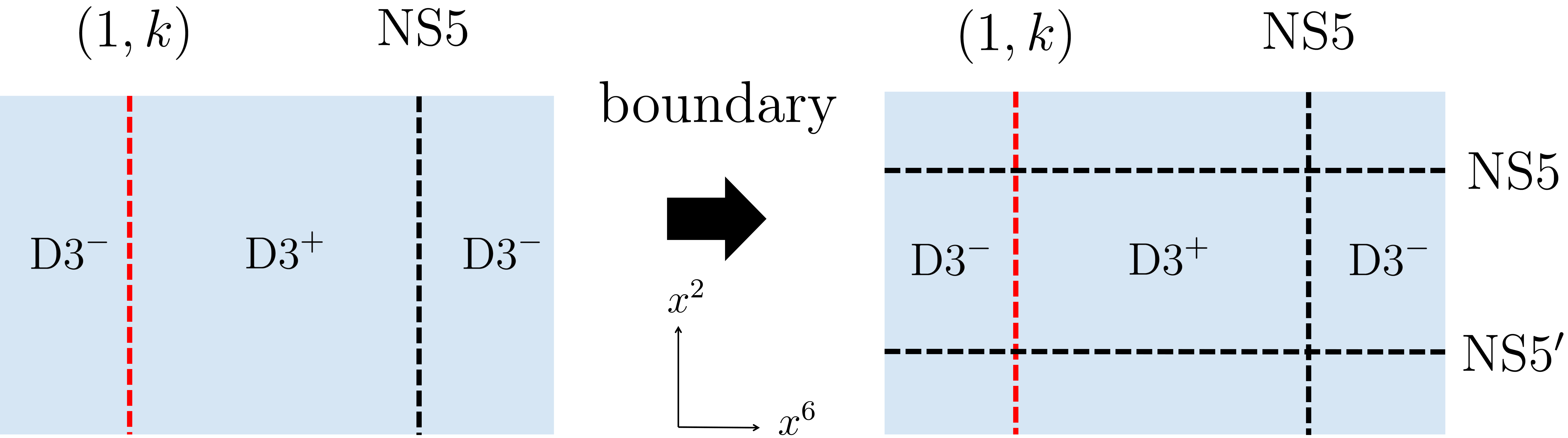}
\caption{The type IIB configuration of the ABJM model with two boundaries of M5- and M5$'$-branes which 
correspond to NS5- and NS5$'$-branes. 
The D3-branes are in finite boxes in the $(x^{2}, x^{6})$-plane. 
}
\label{figabjm2}
\end{center}
\end{figure}

However, this is not the only way to introduce branes corresponding to the
M5-branes in the type IIB configuration. We can alternatively map the M5- and M5$'$-branes 
to NS5- and NS5$'$-branes
in type IIB.
Preserving the same supersymmetry, we can instead add the following NS5-branes 
(see Figure \ref{figabjm2})
\begin{align}
\label{NS5NS5p_2}
\begin{array}{ccccccccccc}
&0&1&2&3&4&5&6&7&8&9\\
\textrm{NS5}
&\circ&\circ& & \nearrow & \nearrow & & \circ &\nearrow& \nearrow& \circ \\
\textrm{NS5$'$}
&\circ&\circ& & \searrow & \searrow & &\circ &\searrow& \searrow & \circ
\end{array}
\end{align}
Now the map to eleven dimensions will results in the following M5-branes
\begin{align}
\label{NSM5M5p_2}
\begin{array}{cccccccccccc}
&0&1&2&3&4&5&6&7&8&9&10\\
\textrm{M5}
&\circ&\circ& & \nearrow & \nearrow & & \circ &\nearrow& \nearrow& \circ & \\
\textrm{M5$'$}
&\circ&\circ& & \searrow & \searrow & &\circ &\searrow& \searrow & \circ &
\end{array}
\end{align}

In either of these cases (\ref{DM5M5p_2}) or (\ref{NSM5M5p_2}) we end up with
M5- and M5$'$-branes which: share the $(x^{0}, x^{1})$ directions with the M2-branes; are
at fixed $x^2$ so can provide a boundary for the M2-branes; and in the
transverse space to the M2-branes, the M5- and M5$'$-branes share two directions
and are orthogonal in the remaining space. Therefore, by simply changing
coordinates in eleven dimensions we can arrive at the brane configuration
(\ref{M2M5M5p}). Note also that in either type IIB configuration, after
introducing either D5- and D5$'$- or NS5- and NS5$'$-branes, the D3-branes can
no longer move in the $(x^{3},x^{4})$ directions.

In the type IIB configuration, we will choose the case where the M5- and
M5$'$-branes are NS5- and NS5$'$-branes. The reason for this is that the
boundary conditions for D3-branes ending on NS5-branes allow preservation of
the full gauge symmetry, and in our M-theory configuration we took boundary
conditions for the M2-branes so that the full gauge symmetry of the Chern-Simons
theory could be preserved \cite{Chung:2016pgt}. 

Now that we have a type IIB configuration, we can consider generalizations of
the M2-M5 system. In particular we could have M2-branes ending on both sides of an M5-brane, and we could also consider more M5- or M5$'$-branes with M2-branes
stretched between them. In type IIB this would correspond to including D3-branes on both sides of the NS5- and NS5$'$-brane, and more generally including several such NS-branes. The advantage of the IIB configuration is that it is possible
to describe the field theory on the D3-branes in terms of open strings. Mapping
this back to M-theory should indicate the effect of having two ABJM theories
coupled through the brane configuration of M2-branes ending on both sides of an M5-brane. Some results in this direction have been derived by
Niarchos \cite{Niarchos:2015lla}, without M5$'$-branes or topological twisting.
It would be interesting to understand the relation in detail.

We leave a full analysis of the type IIB configurations to future work. However,
we note that our expectation is that the configuration with $N$ D3$^{+}$- and
$N$ D3$^{-}$-branes stretched between an NS5- and an NS5$'$-brane gives a
$GL(N \vert N)$ WZW model after taking the limit of coincident NS5- and
NS5$'$-branes and dualizing to M-theory. If we introduce a stack of $M$
D3$^{+}$- and $M$ D3$^{-}$-branes on the other side of the NS5$'$-brane
and allow these to end on an addition NS5- or NS5$'$-branes, we
will arrive at a $GL(N \vert N) \times GL(M \vert M)$ WZW model with
bi-fundamental matter from the open strings connecting the D3-branes across the
NS5$'$-brane. In M-theory this would correspond to the configuration with (along
increasing $x^2$) M5 - $N$ M2 - M5$'$ - $M$ M2 - M5. While we hope to return to
this type IIB description in future, for this paper we now focus on the case
with just the single stacks of $N$ D3$^{+}$- and D3$^{-}$-branes.

\subsection{Supergroup WZW model}
\label{subsecsgwzw}
The action of the supergroup WZW model for maps $s:\Sigma\rightarrow SG$ 
from a two-dimensional Euclidean Riemann surface $\Sigma$ to the supergroup $SG$ is given by 
\begin{align}
\label{sgwzwlag}
S[s]&=
-\frac{k}{8\pi}
\int_{\Sigma} d^{2}x 
\left( s^{-1}\partial^{\alpha}s, s^{-1}\partial_{\alpha}s 
\right)
-\frac{ik}{24\pi}\int_{M} d^{3}x \epsilon^{\mu\nu\lambda}
\left( s^{-1}\partial_{\mu}s, 
[s^{-1}\partial_{\nu}\lambda, s^{-1}\partial_{\lambda}s]\right)
\end{align}
where $k\in \mathbb{Z}$ is the level 
\footnote{
As in \cite{Okazaki:2015fiq} our description of the M2-M5 system is for the
case $k=1$, for which the BLG and ABJM models describe flat membranes.
}. 
Here the second term is the WZ term integrated over a three-manifold $M$ 
whose boundary is $\Sigma$.

The action (\ref{sgwzwlag}) is invariant under the transformation
\begin{align}
\label{sym1a1}
s(z,\overline{z})\rightarrow \Omega(z) s(z,\overline{z}) 
\overline{\Omega}^{-1}(\overline{z})
\end{align}
where $\Omega(z)$ and $\overline{\Omega}(\overline{z})$ 
are arbitrary $SG$-valued functions of the complex variables 
$z=x^{0}+ix^{1}$ and $\overline{z}=x^{0}-ix^{1}$. 
This realizes the semi-local symmetry $SG(z)\times SG(\overline{z})$, 
the direct product symmetry group of left and right multiplications. 
Under the infinitesimal transformation
$\Omega(z)=\mathbb{I}+\omega(z)$, 
$s$ transforms as $\delta_{\omega}s=\omega s$ 
and the action (\ref{sgwzwlag}) is invariant. 
Hence we find the conserved currents
\begin{align}
\label{curr1a1}
J(z)&
=J^{a}(z)T_{a}
=-k\partial_{z}s\cdot s^{-1}
\end{align}
where $T^{a}$ is a generator of $\mathfrak{sg}$. 
The conservation of the currents can be derived from the classical equations of motion $\partial_{\overline{z}}J=0$, which ensure that $J$ is holomorphic. 
Let us concentrate only on the holomorphic current $J$.  
Substituting the transformation 
$\delta_{\omega}J$ into the Ward identity, we obtain the OPE
\begin{align}
\label{curr2a1}
J^{a}(z)J^{b}(w)&\sim
\frac{k(T^{a},T^{b})}{(z-w)^{2}}
+\frac{ [T^{a},T^{b}]_{c} J^{c}(w)}{z-w}.
\end{align}
Since the current is an analytic function of $z$, 
it can be expanded as 
\begin{align}
\label{curr2a2}
J^{a}(z)&=\sum_{n=-\infty}^{\infty}
\frac{J_{n}^{a}}{z^{n+1}}.
\end{align}
Then the OPE (\ref{curr2a1}) leads to the affine Lie superalgebra $\widehat{\mathfrak{sg}}$
\begin{align}
\label{curr2a3}
[J_{n}^{a},J_{m}^{b}]&=
[T^{a},T^{b}]_{c}J_{n+m}^{c}
+m(T^{a},T^{b})\delta_{n+m,0} k.
\end{align}

\section{Affine Lie Superalgebra}
\label{secaff}
Due to the underlying symmetry algebra (\ref{curr2a3}), 
we need to study the affine Lie superalgebra
\begin{align}
\label{aff1a1}
\widehat{\mathfrak{sg}}&=
(\mathbb{C}[t,t^{-1}]\otimes
\mathfrak{sg})
\oplus \mathbb{C}K\oplus \mathbb{C}d .
\end{align}
Here $\mathbb{C}[t,t^{-1}]\otimes
\mathfrak{sg}\oplus \mathbb{C}K$ 
is a central extension of the loop algebra 
$\widetilde{\mathfrak{sg}}=\mathbb{C}[t,t^{-1}]\otimes \mathfrak{sg}$ 
with $\mathbb{C}[t,t^{-1}]$ being Laurent polynomial in variable $t$,
$K$ being a central element called the level, and 
$d=t\frac{d}{dt}$ being the derivation. 
The generators of $\widehat{\mathfrak{sg}}$ obey the commutation relations 
\begin{align}
\label{aff1a2a}
[at^{n},bt^{m}]&=[a,b]t^{n+m}+m\delta_{m+n,0}( a,b ) K,\\
\label{aff1a2b}
[d,at^{n}]&=n a t^{n},& [K,\widehat{\mathfrak{sg}}]&=0,
\end{align}
and the non-degenerate supersymmetric invariant bilinear form is 
\begin{align}
\label{affbi1}
( at^{n}, bt^{m} )&=\delta_{m+n,0}( a,b ),&
(\widetilde{\mathfrak{sg}}, \mathbb{C}K+\mathbb{C}d)&=0, \\
\label{affbi2}
( K,K )&=( d,d )=0,&
( K,d )&=1,
\end{align}
with $a,b\in \mathfrak{sg}$, $m,n\in \mathbb{Z}$. 
Note that in the physical setup (\ref{curr2a3}) 
the derivation $d$ corresponds to the Virasoro generator $-L_{0}$  
while the level $K$ is the constant value for the 
$SG$ WZW model (\ref{sgwzwlag}).

The Cartan subalgebra $\widehat{\mathfrak{h}}$ of $\widehat{\mathfrak{sg}}$
can be defined in terms of a Cartan subalgebra $\mathfrak{h}$ of the finite
Lie superalgebra $\mathfrak{sg}$,
\begin{align}
\label{affcar1a}
\widehat{\mathfrak{h}}&=
\mathfrak{h}+\mathbb{C}d+\mathbb{C}K .
\end{align}
We will introduce the coordinate on $\widehat{\mathfrak{h}}$ 
\begin{align}
\label{affcar1b}
h&:=2\pi i(-\tau d+z+tK)
\end{align}
with $\tau$, $t\in \mathbb{C}$, $z\in \mathfrak{h}$.

\subsection{Roots and weights}
\label{subsecroot}
The non-degenerate bilinear form of $\widehat{\mathfrak{sg}}$ is extended to 
$\widehat{\mathfrak{h}}$ as $( \mathfrak{h},\mathbb{C}K+\mathbb{C}d )=0$ 
and one gets the dual $\widehat{\mathfrak{h}}^{*}$ of $\widehat{\mathfrak{h}}$. 
The roots and weights belong to the dual $\widehat{\mathfrak{h}}^{*}$ of $\widehat{\mathfrak{h}}$. 
The root space is 
\begin{align}
\label{affroot1a}
\widehat{\mathfrak{h}}^{*}&=
\mathfrak{h}^{*}\oplus \mathbb{C}\delta \oplus \mathbb{C}\Lambda_{0}
\end{align}
where the elements 
$\delta$ and $\Lambda_{0}$ of $\widehat{\mathfrak{h}}^{*}$ are defined by 
\begin{align}
\label{affroot1a1}
\delta|_{\mathfrak{h}+\mathbb{C}K}&=0,& 
\delta(d)&=1, \\
\label{affroot1a2}
\Lambda_{0}|_{\mathfrak{h}+\mathbb{C}d}&=0,&
\Lambda_{0}(K)&=1.
\end{align}
and they are represented by 
$\delta=(0,0,1)$ and $\Lambda_{0}=(1,0,0)$.
As $\widehat{\mathfrak{h}}$ is identified with $\widehat{\mathfrak{h}}^{*}$ 
by the bilinear form (\ref{affbi2}), 
we have 
\begin{align}
\label{affroot1c}
\delta&=K, &
\Lambda_{0}&=d. 
\end{align}
Let 
$\widehat{\Delta}\subset \widehat{\mathfrak{h}}^{*}, 
\widehat{\Delta}_{\overline{0}}$ and $\widehat{\Delta}_{\overline{1}}$ 
be the set of roots, the subset of even and odd roots respectively. 
$\widehat{\Delta}_{\overline{0}}$ turns out to be a union of 
a finite number of root systems 
$\widehat{\Delta}^{\textrm{re}+}:=\{ \alpha+s\delta|\alpha\in \Delta, s>0 \} \cup \Delta^{+}$ 
of the affine Lie superalgebra 
with the same primitive imaginary roots 
$\widehat{\Delta}^{\textrm{im}+}:=\{ s\delta | s>0 \}$. 
We define a coroot as $\alpha^{\vee}=\frac{2\alpha}{(\alpha,\alpha)}$ for 
non-isotropic root $\alpha\in \widehat{\Delta}$ and 
$\alpha^{\vee}=\alpha$ for 
isotropic root $\alpha\in \widehat{\Delta}$.

The set of simple roots of $\widehat{\mathfrak{sg}}$ is given by 
$\widehat{\Pi}=\Pi\cup \alpha_{0}$,  
where $\Pi=(0,\alpha_{i},0)$ with $\alpha_{i}$ being simple roots of $\mathfrak{sg}$ 
and $\alpha_{0}:=\delta-\theta = (0,-\theta,1)$ with $\theta$ being the highest root of $\mathfrak{sg}$,  
which is defined by $\theta=\sum_{i=1}^{N+M-1} k_{i}\alpha_{i}\in \Delta_{+}$ 
so that $\sum_{i=1}^{N+M-1}k_{i}$ is maximal for $\mathfrak{gl}(N|M)$.
For example, the sets of simple roots of $\widehat{\mathfrak{gl}}(N|N)$ 
which consist of isotropic roots are 
\begin{align}
\label{sroot1a}
\left\{
\delta-\epsilon_{1}-\delta_{N}, 
\epsilon_{1}-\delta_{1}, 
\delta_{1}-\epsilon_{2}, 
\cdots, 
\delta_{N-1}-\epsilon_{N}, 
\epsilon_{N}-\delta_{N}
\right\}. 
\end{align}
%
The Borel subalgebra $\widehat{\mathfrak{b}}$ of $\widehat{\mathfrak{sg}}$ is given by 
\begin{align}
\label{bor1}
\widehat{\mathfrak{b}}
&=\widehat{\mathfrak{h}}\oplus \widehat{\mathfrak{n}}^{+}=\widehat{\mathfrak{h}}\oplus \mathfrak{n}^{+}\oplus 
\left(
\bigoplus_{n>0} t^{n}\otimes \mathfrak{sg}
\right).
\end{align}

%
%

%
A weight $\Lambda\in \widehat{\mathfrak{h}}^{*}$ takes the form 
$(k,\lambda,n)$ where $\lambda$ is the weight of $\mathfrak{sg}$. 
The fundamental weight $\Lambda_{i}\in \widehat{\mathfrak{h}}^{*}$ is defined by 
\begin{align}
\label{fweight1}
(\Lambda_{i},\alpha_{j}^{\vee})&=\delta_{ij},& 
(\Lambda_{i},d)&=0 
\end{align}
and the label of the weight $\lambda$ by 
\begin{align}
\label{fweight2}
m_{i}&=(\Lambda,\alpha_{i}^{\vee}). 
\end{align}

\subsection{Weyl group}
\label{subsecweyl}
The affine Weyl vector $\widehat{\rho}$ is defined by
\begin{align}
\label{aweyl1}
\widehat{\rho}&=\rho+h^{\vee}\Lambda_{0}.
\end{align}
It obeys $\left( \widehat{\rho},\alpha\right )=\frac12 \left( \alpha,\alpha\right )$ for $\forall \alpha\in \widehat{\Pi}$, 
$\left ( \widehat{\rho},d\right )=0$ and $\left ( \widehat{\rho},K\right )=h^{\vee}$. 
For $\alpha\in \mathfrak{h}^{*}$ 
we define $t_{\alpha}\in \textrm{Aut}(\hat{\mathfrak{h}}^{*})$ by 
\begin{align}
\label{aweyl2}
t_{\alpha}(\Lambda)
&=\Lambda+\Lambda(K)\alpha
-\left(
(\Lambda,\alpha)+\frac12 (\alpha,\alpha)\Lambda(K)
\right)\delta. 
\end{align}
The affine Weyl group is 
\begin{align}
\label{aweyl3}
\widehat{W}&=
W\ltimes 
\left\{
t_{\alpha}|\alpha\in L
\right\}
\end{align}
where $W$ is the Weyl group of $\mathfrak{sg}$ 
and $L\subset \mathfrak{h}$ is the coroot lattice.

\subsection{Representations}
\label{subsecmodule}
For each weight $\Lambda\in \widehat{\mathfrak{h}}^{*}$ 
one can define the irreducible highest weight module $L(\Lambda)$ 
over $\widehat{\mathfrak{sg}}$ such that there exists a non-zero vector $v_{\Lambda}$ satisfying
\begin{align}
\label{affw1a1}
hv_{\Lambda}&=\Lambda(h)v_{\Lambda},& &\textrm{for $h \in \widehat{\mathfrak{h}}$}, \\
\label{affw1a2}
\mathfrak{n}^{+}v_{\Lambda}&=0, \\
\label{affw2a3}
\left(t^{n}\otimes \mathfrak{sg}\right) v_{\Lambda}&=0,& &\textrm{for $n>0$}.
\end{align}
The central element $K$ on $L(\Lambda)$ is the scalar $k=\Lambda(K)$ 
called the level in (\ref{affroot1a}). 
The irreducible highest weight module $L(\Lambda)$ is called integrable if 
(i) $\dim L(\Lambda)<\infty$ 
and (ii) $t^{n}\otimes \mathfrak{sg}_{\alpha}$ are locally nilpotent for all $\alpha\in \Delta_{\overline{0}}^{\sharp}$ and $n\in \mathbb{Z}$. 
It is known that 
$L(\Lambda)$ is integrable if the number $\frac{2(\Lambda,\alpha)}{(\alpha,\alpha)}$ and 
$\frac{2(\Lambda,K-\theta)}{(\theta,\theta)}$ are non-negative integers 
for all simple roots $\alpha\in \widehat{\Pi}$ and the highest root $\theta$. 
The necessary condition of integrability of $L(\Lambda)$ over $\mathfrak{gl}(N|M)$ is \cite{MR1810948}
\begin{align}
\label{direp1}
m_{i}&\in \mathbb{Z}_{+},& 
m'&=m_{0}+m_{N}-\sum_{i=N+1}^{N+M-1}m_{i}\in \mathbb{Z}_{+}
\end{align}
and the sufficient condition is \cite{MR1810948}
\begin{align}
\label{direp2}
m'&\ge M
\end{align}
for $N\ge 2$.

%
Let $S$ be a subset of a simple root system $\Pi$. 
We call it a $(\lambda+\rho)$-maximal isotropic subset 
if it consists of $d$ pairwise orthogonal isotropic roots 
$\{\beta_{i}\}$, $i=1,\cdots,d$ that are also orthogonal 
to $\lambda+\rho$, i.e. \cite{MR1327543,MR3200431}
\begin{align}
\label{tame1}
\left( \lambda+\rho,\beta_{i}\right )&=0,& 
\left( 
\beta_{i},\beta_{j}
\right)&=0. 
\end{align}
The number $d$ of linearly independent pairwise orthogonal isotropic roots is
called the atypicality of $L(\lambda)$. 
The atypicality of a simple finite dimensional module does not depend on the choice of simple root system 
and the maximal number $d$ of the Lie superalgebra $\mathfrak{sg}$ is called
the defect and denoted by $\mathrm{def}(\mathfrak{sg})$.

An irreducible highest weight module $L(\lambda)$ over $\mathfrak{sg}$ is called typical  
if $S$ is empty and atypical or tame otherwise. 
Similarly, an irreducible highest weight module $L(\Lambda)$ of level $K$ over $\widehat{\mathfrak{sg}}$ is called atypical or tame 
if the corresponding module $L(\lambda)$ over the finite part $\mathfrak{sg}$ of $\widehat{\mathfrak{sg}}$ is atypical 
and if $K+h^{\vee}\neq 0$ \cite{MR1327543, MR3200431}. 

Note that the irreducible highest weight module $L(\lambda)$ 
is characterized by the vectors annihilated by $\mathfrak{n}^{+}$ 
acting as the raising operators.
However, the choice of $\mathfrak{n}^{+}$ is not unique 
but depends on the Weyl group $W$ that permutes the different weights. 
To characterize $L(\lambda)$ over $\mathfrak{sg}$ so that 
the choice of $\mathfrak{n}^{+}$ does not depend on $W$, 
we need to take the shifted weight $\lambda+\rho$ on which $w\in W$ acts.

\section{Branes and Weight Diagram}
\label{secdiagram}

\subsection{Weight diagram}
\label{subsecw1}
In terms of the basis $\{\epsilon_{1}, \cdots, \epsilon_{N}; \delta_{1},\cdots, \delta_{M}\}$ of $\mathfrak{h}^{*}$, 
one can write the dominant integrable weight $\lambda$ of the irreducible highest weight modules $L(\lambda)$ as 
\begin{align}
\label{wdia1}
\lambda+\rho&=
\sum_{i=1}^{N}x_{i}\epsilon_{i}
-\sum_{k=1}^{M}y_{k}\delta_{k}
\end{align}
where the integral condition requires that the coefficients $x_{i}$ and $y_{k}$ are integers 
and the dominant condition is satisfied by the ordering $x_{1}\ge \cdots \ge x_{N}$, $y_{1}\le \cdots \le y_{M}$. 
It can be represented diagrammatically in terms of the weight diagram,
and the irreducible characters over the Lie superalgebras 
have been computed using a combinatorial algorithm
\cite{MR2918294, MR2734963, MR2928846, MR3397407}. 
Consider a horizontal number line with vertices labelled by 
a set of consecutive integers $n$ in increasing order from left to right. 
Then we label the vertex of $n$ by 
\begin{align}
\label{wdia2}
\begin{cases}
\vee& \textrm{if $n \in \{x_{i}\} \cap \{y_{k}\}$} \cr
>&\textrm{if $n \in \{x_{i}\} \setminus \{y_{k} \}$} \cr
<&\textrm{if $n \in \{y_{k}\}\setminus \{x_{i}\}$ } \cr
\wedge&\textrm{if $n \notin \{x_{i}\} \cup \{y_{k}\}$}. \cr
\end{cases}
\end{align}
Each $\vee$ corresponds to an atypical root $\beta$ and 
the degree $d$ of atypicality of $\lambda$ is the number of $\vee$'s in the weight diagram. 
The dominant weight is uniquely determined by the weight diagram. 

For example, the weight 
\begin{align}
\label{wdia1a}
\lambda+\rho&
=9\epsilon_{1}+5\epsilon_{2}+3\epsilon_{3}+2\epsilon_{4}
-\delta_{1}-3\delta_{2}-7\delta_{3}-9\delta_{4}
\end{align}
corresponds to the following weight diagram
\begin{align}
\label{wdia1b}
\begin{array}{cccccccccccc}
\wedge&\wedge&<&>&\vee&\wedge&>&\wedge&<&\wedge&\vee&\wedge \\
-1&0&1&2&3&4&5&6&7&8&9&10 \\
\end{array}.
\end{align}
The $\lambda+\rho$-maximal isotropic subset is 
\begin{align}
\label{wdia1c}
S&=\left\{
\epsilon_{1}-\delta_{4},\epsilon_{3}-\delta_{2}
\right\}
\end{align}
and the atypicality of the corresponding irreducible highest weight module $L(\Lambda)$ is $d=2$.

One can consider certain combinatorial operations on the weight diagrams 
by moving $\vee$'s and $\wedge$'s at specific positions to other locations 
\cite{MR2918294, MR2734963, MR2928846}. 
We define a right move $R_{i\rightarrow j}(\lambda)$ 
on the weight diagram $\lambda$ by 
exchanging (counting from the left) the $i$-th $\vee$ with a
$\wedge$ to its right. This $\wedge$ is specified
in such a way that there are exactly $k \equiv j -i$ $\vee$'s and the same number of $\wedge$'s
between the $i$-th $\vee$ and this $\wedge$.
As a consequence, the $i$-th $\vee$ moves to become the $j=(i+k)$-th $\vee$.
For example, 
for the weight diagram (\ref{wdia1b}) 
$R_{1\rightarrow 2}\circ 
R_{1\rightarrow 1}\circ 
R_{1\rightarrow 1}\circ 
R_{1\rightarrow 1} (\lambda)$ is 
\begin{align}
\label{wdia2a}
\begin{array}{ccccccccccccc}
\wedge&\wedge&<&>&\wedge&\wedge&>&\wedge&<&\wedge&\vee&\wedge&\vee \\
-1&0&1&2&3&4&5&6&7&8&9&10&11 \\
\end{array}.
\end{align}
Note for the last step that all locations to the right (or left)
of the weight diagram are filled by $\wedge$'s.
The right move $R_{i\rightarrow j}$ corresponds to a raising operator for the
corresponding module \cite{MR2928846}. 

A left move $L_{i\leftarrow j}$ is similarly defined 
by swapping (still counting from the left) the $j$-th $\vee$ with a $\wedge$ 
to its left, again separated by $k \equiv j-i$ $\vee$'s and $k$
$\wedge$'s. Then the $j$-th $\vee$ is shifted to the $i=(j-k)$-th $\vee$.
For example, for the weight diagram (\ref{wdia1b}) 
$L_{1\leftarrow 2}\circ L_{2\leftarrow 2}\circ L_{2\leftarrow 2}\circ
L_{2\leftarrow 2} (\lambda)$ gives
\begin{align}
\label{wdia2b}
\begin{array}{ccccccccccccc}
 \vee&\wedge&\wedge&<&>&\vee&\wedge&>&\wedge&<&\wedge&\wedge&\wedge \\
-2&-1&0&1&2&3&4&5&6&7&8&9&10 \\
\end{array}.
\end{align}
This operation corresponds to a lowering operator in the corresponding
module \cite{MR2928846}.

\subsection{Brane construction}
\label{subsecw2}
Now we return to the $GL(N|N)$ WZW model describing the M2-M5 brane system. 
We argue that the dominant integrable weight $\lambda$ 
of the irreducible highest weight atypical module $L(\Lambda)$ over $\widehat{\mathfrak{gl}}(N|N)$ 
corresponds to the vacuum configuration of branes. 

Let $C$ (resp.\ $C'$) be the M-theory 3-form `$C$-field' on the M5-brane (resp.\ M5$'$-brane) 
and let $\Sigma_{a}$, $a=1,\cdots,N$ be the 2-cycle wrapped by $a$-th M2-brane. 
In the two dimensional intersection with the M5-brane (resp.\ M5$'$-brane) $\Sigma_{a}$, 
Abelian gauge fields $\{A^{i}\}$, $i=1, \cdots, N$ (resp.\ $\{{A'}^{k}\}$, $k=1, \cdots, N$) 
arise from the Kaluza-Klein reduction of the M-theory 3-form
\begin{align}
\label{c3field}
C&=\sum_{i=1}^{N}A^{i}\wedge \Sigma_{i},& 
C'&=\sum_{k}^{N}{A'}^{k}\wedge \Sigma_{k}. 
\end{align}
The presence of the M5- and M5$'$-branes independently carrying 
$N$ M2-brane charges of the $C$-field implies that 
one can specify data of the M2-M5 system 
by a choice of two sets of vector bundles $E\rightarrow \Sigma_{i}$, $E'\rightarrow \Sigma_{k}$ 
and connections on $E$, $E'$.
From the M2-brane point of view they are viewed as global charges. 
We denote the eigenvalue of the $i$-th M2-brane charge in the M5-brane
by $x_{i}\in \mathbb{Z}$, $i=1,\cdots, N$ 
and that of the $k$-th M2-brane charge for the M5$'$-brane 
by $y_{k}\in \mathbb{Z}$, $k=1,\cdots, N$ respectively. 
Then we can obtain a unique weight diagram from the brane configuration 
by considering an integer coordinate 
and putting a symbols $\{\vee, >, <, \wedge \}$ on it in the same manner as (\ref{wdia2}). 

Similarly Mikhaylov and Witten \cite{Mikhaylov:2014aoa} 
point out that a vacuum configuration of the brane system with $N$ D3-branes
ending on one side, and $M$ D3-branes ending on the other side, of a single
NS5-brane corresponds to the dominant integrable weight $\lambda$ 
of $\mathfrak{u}(N|M)$ and its weight diagram (see Figure \ref{mw2}). 
In that case the two sequences $\{x_{i}\}$ and $\{y_{k}\}$ would represent the charges 
of wrapped D3-branes under the RR fields.

This construction gives interesting physical implications of the weight diagram. 
The non-zero eigenvalues of M2-brane charge correspond to $\vee$'s  
that are shared by both M5-branes,
and to $>$ or $<$ that are taken by only one of the M5-branes. 
Since the limit in which the separation of the M5-branes is taken to zero 
require the same eigenvalues for both M5-branes, 
the $\vee$'s are identified with the M2-branes which are suspended between the M5- and M5$'$-brane. 
Thus the atypicality, that is the number of $\vee$'s, 
is the number of M2-branes attached to both M5-branes. In particular, for
$GL(N \vert N)$ arising from $N$ M2-branes all stretched between the two M5-branes, the modules of interest have maximal atypicality $N$.

\begin{figure}
\begin{center}
\includegraphics[width=8.5cm]{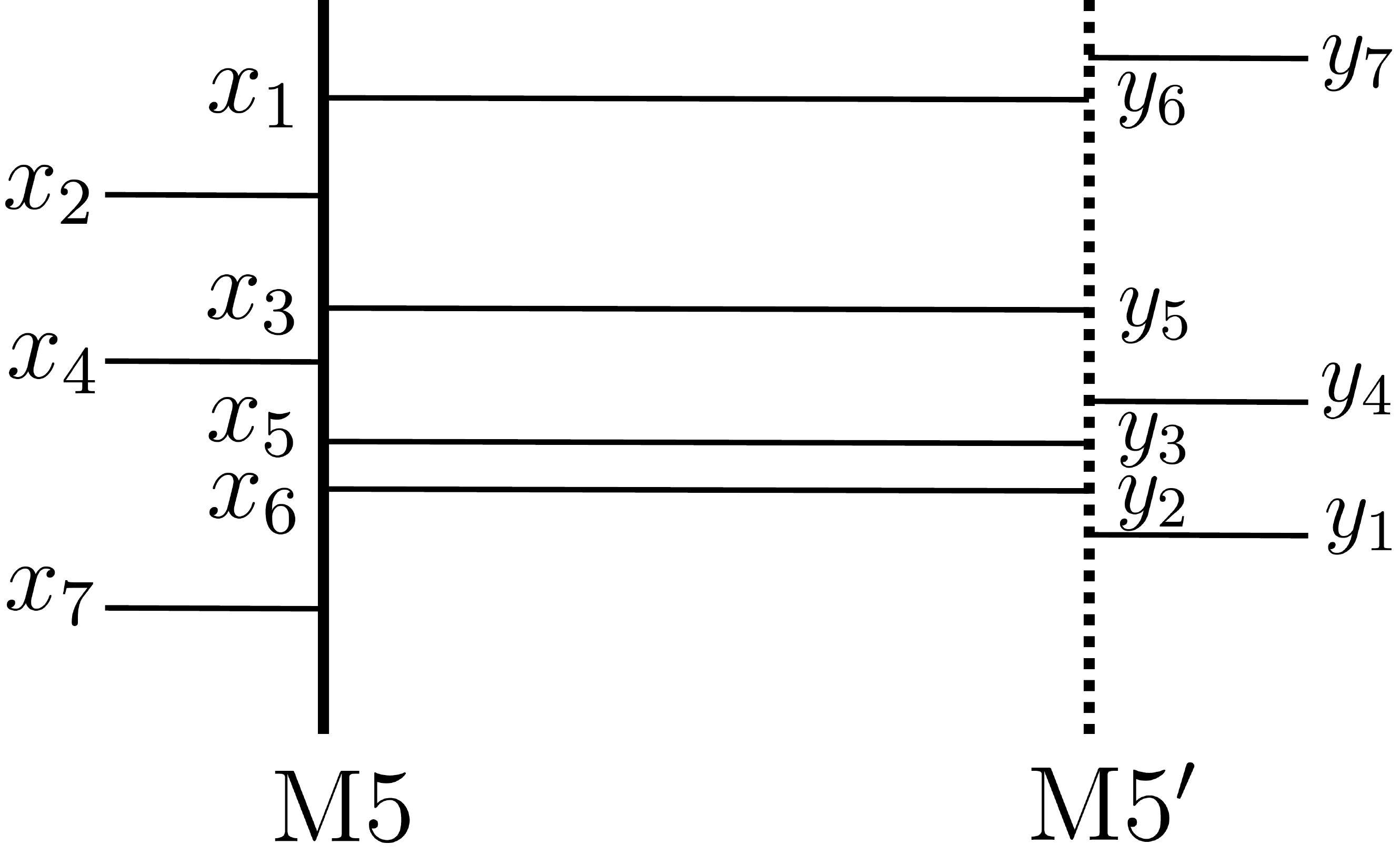}
\caption{$(N-d)$ M2-branes attached to one of the M5-branes and 
$d$ M2-branes stretched between the two M5-branes. 
The vertical bold (resp.\ dotted) line represent M5- (resp.\ M5$'$-)brane, 
the horizontal lines denotes M2-branes in the $x^{2}$ direction. 
The M2-brane charges $\{x_{i}\}$ and $\{y_{k}\}$ are illustrated as the heights 
of the M2-branes.
Here is the case with $N=7$ and $d=4$. }
\label{mdiagram1}
\end{center}
\end{figure}

\begin{figure}
\centering
    \begin{tabular}{cc}
          \includegraphics[scale=0.275]{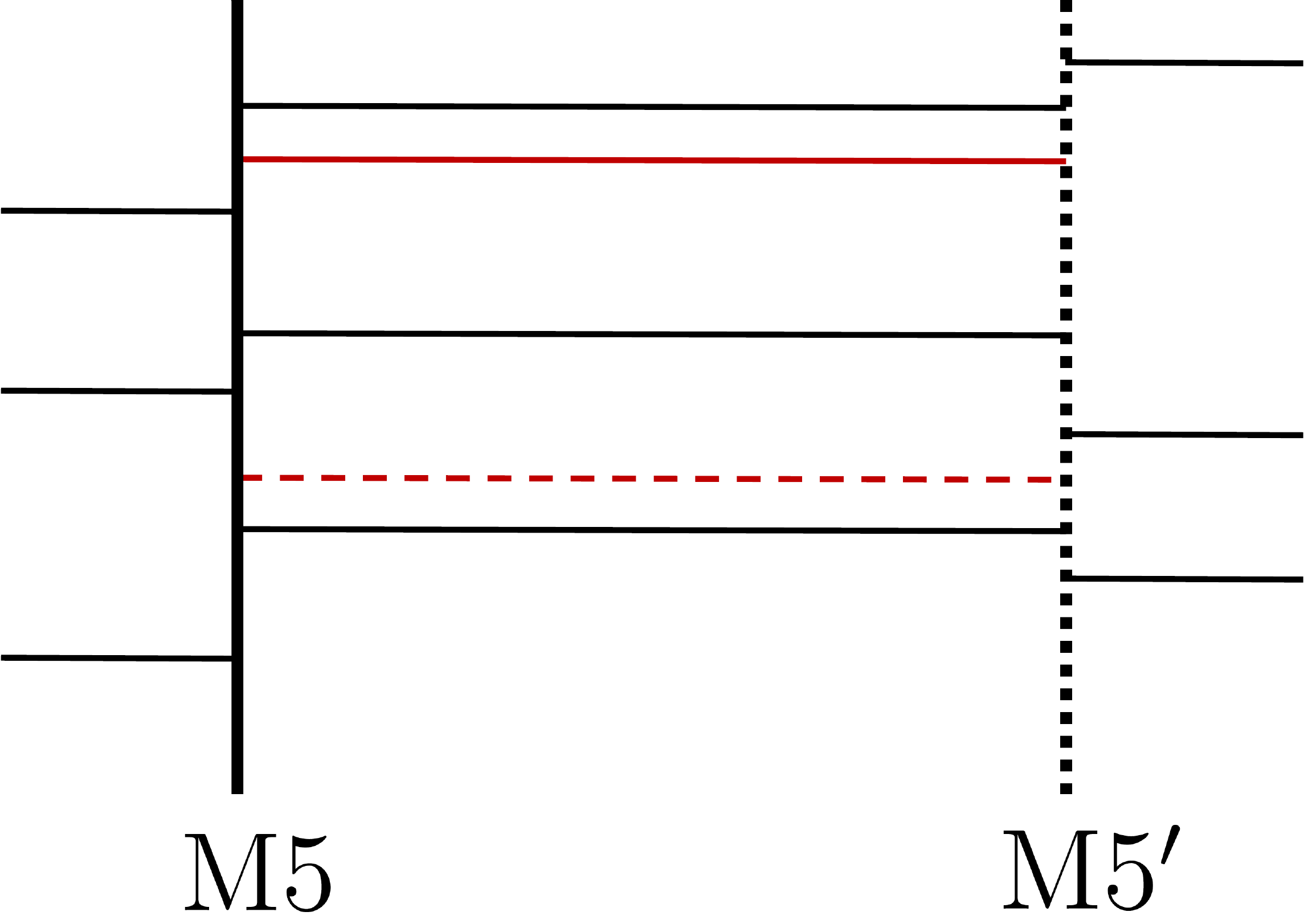}&\includegraphics[scale=0.275]{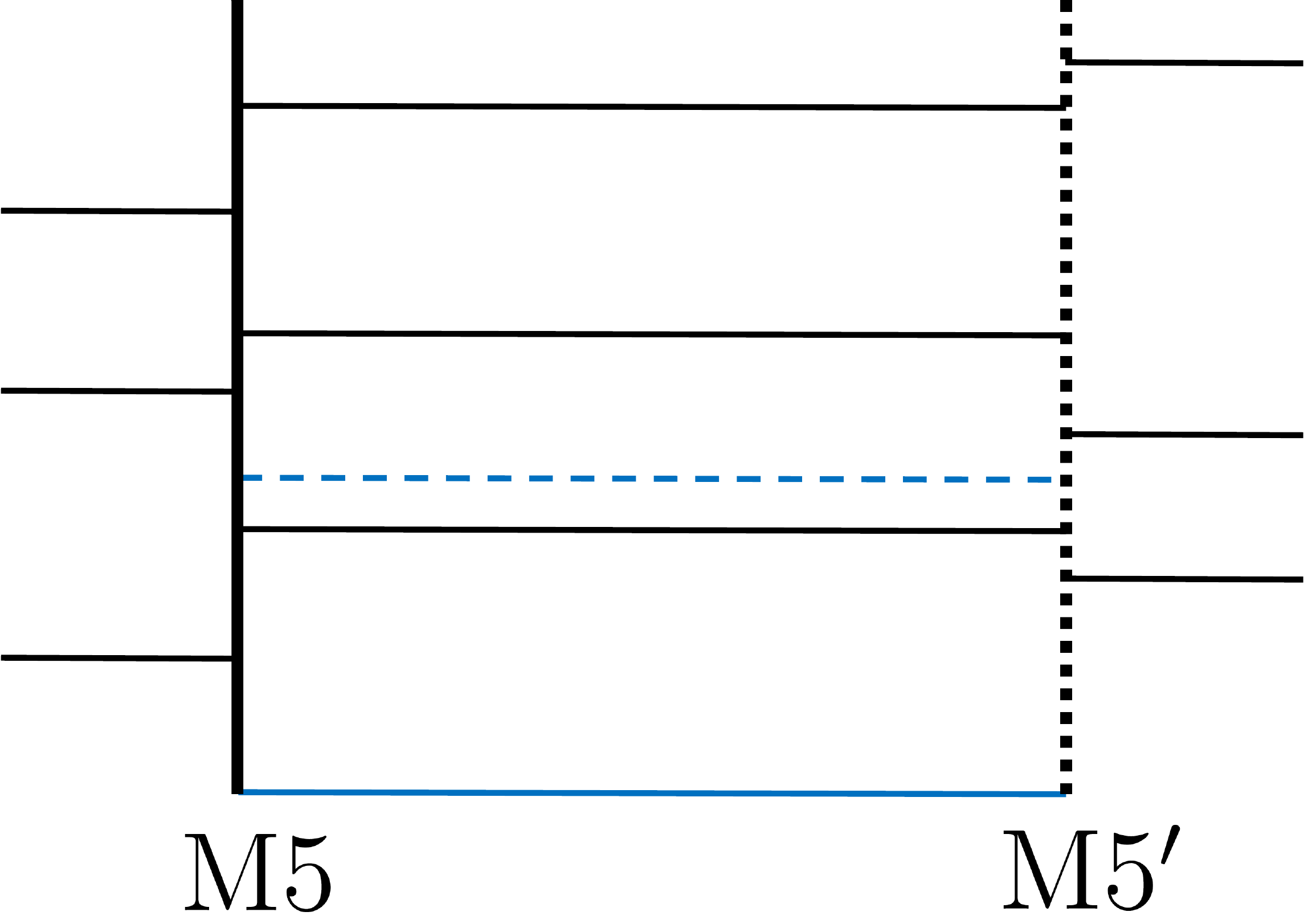}\\
          (a) The right move $R_{2\rightarrow3}$&(b) The left move
	 $L_{1\leftarrow 2}$
\end{tabular}
    \caption{The operation of the right move and the left move on the brane configuration in Figure \ref{mdiagram1}. 
    The right move $R_{2\rightarrow 3}$ lifts the second mode at $5$ (shown in dotted red) to
 the third at $11$ (shown in red) while 
    the left move $L_{1\leftarrow 2}$ reduces the second mode of the
 location $5$ (shown in dotted blue) to $0$ (shown in blue).}
    \label{rldia1}
\end{figure}

For example, consider the brane configuration in Figure \ref{mdiagram1} 
with $d=4$ M2-branes stretched between M5- and M5$'$-brane  
and $(N-d)=(7-4)=3$ M2-branes attached to one of them. 
Set the eigenvalues of the $i$-th M2-brane charge for the M5-brane as 
$\{x_{i}\}=\{12, 10, 8, 7, 5, 4, 1\}$ 
and those of the $k$-th M2-brane charge for the M5$'$-brane as 
$\{y_{k}\}=\{3, 4, 5, 6, 8, 12, 13\}$, 
which correspond to the heights of the M2-branes in Figure \ref{mdiagram1}.
Then the corresponding weight reads
\begin{align}
\label{bdia1}
\begin{array}{ccccccccccccccc}
\wedge&\wedge&>&\wedge&<&\vee&\vee&<&>&\vee&\wedge&>&\wedge&\vee&< \\
-1&0&1&2&3&4&5&6&7&8&9&10&11&12&13 \\
\end{array}
\end{align}
and the dominant weight is 
\begin{align}
\label{bdia2}
\lambda+\rho&=
12\epsilon_{1}+10\epsilon_{2}+8\epsilon_{3}+7\epsilon_{4}+5\epsilon_{5}+4\epsilon_{6}+\epsilon_{7}\nonumber\\
&-3\delta_{1}-4\delta_{2}-5\delta_{3}-6\delta_{4}-8\delta_{5}-12\delta_{6}-13\delta_{7}. 
\end{align}
The $\lambda+\rho$ -maximal isotropic subset $S$ is 
\begin{align}
\label{bdia3}
S&=
\{
\epsilon_{1}-\delta_{6}, \epsilon_{3}-\delta_{5}, \epsilon_{5}-\delta_{3}, \epsilon_{6}-\delta_{2}
\}
\end{align}
and the atypicality of the module is $d=4$, 
that is the number of M2-branes stretched between the M5- and M5$'$-brane. 
The right move $R_{2\rightarrow 3}(\lambda)$ gives the weight diagram 
\begin{align}
\label{lrdia1}
\begin{array}{ccccccccccccccc}
\wedge&\wedge&>&\wedge&<&\wedge&\vee&<&>&\vee&\wedge&>&\vee&\vee&< \\
-1&0&1&2&3&4&5&6&7&8&9&10&11&12&13 \\
\end{array}
\end{align}
and the left move $L_{1\leftarrow 2}$ yields the weight diagram
\begin{align}
\label{rdia1}
\begin{array}{ccccccccccccccc}
\wedge&\vee&>&\wedge&<&\vee&\wedge&<&>&\vee&\wedge&>&\wedge&\vee&< \\
-1&0&1&2&3&4&5&6&7&8&9&10&11&12&13 \\
\end{array}.
\end{align}
They correspond to new charge assignments 
of the brane configuration depicted in Figure \ref{rldia1}. 
The right move $R_{i\rightarrow j}$ 
and the left move $L_{i\leftarrow j}$ are respectively interpreted as 
a raising operator and a lowering operator of the suspended M2-brane charges. 

Quantum mechanically, a transition amplitude is given by a weighted sum over all paths as the Feynman path integral. 
As shown in Figure \ref{rldia1}, it will be achieved by summing over all possible paths of excitation modes 
by acting with raising or lowering operators. 
However, it can be now rephrased as a sum over 
all possible paths of the sequence of left-moves, or equivalently right-moves 
with a weight characterized by multiplicity of the path 
\footnote{Interestingly the terminology \textit{path} is also used 
for the collection of the left-moves and right-moves
in the mathematical literature 
\cite{MR2734963, MR2928846}}.
Therefore the dominant weight 
of the irreducible highest weight atypical module $L(\Lambda)$ 
over the underlying symmetry $\widehat{\mathfrak{gl}}(N|N)$ 
can be determined by the vacuum configuration of the M2-M5 system.

\begin{figure}
\begin{center}
\includegraphics[width=12.5cm]{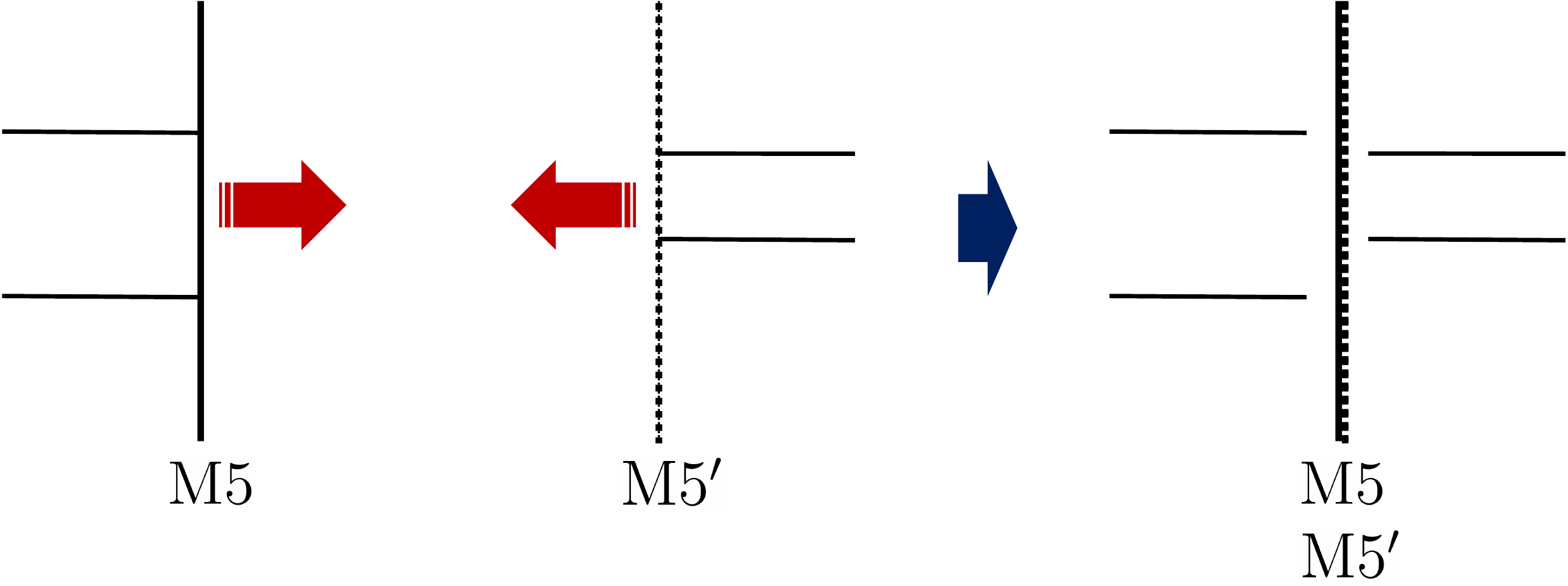}
\caption{The M2-M5 system for typical module with $N=2$, $d=0$, 
for which no M2-brane is stretched between the M5-branes.}
\label{typ1}
\end{center}
\end{figure}

In the absence of atypical roots 
the dominant integral weight $\lambda$ defines a typical highest weight module $L(\lambda)$ \cite{MR0444725}. 
In the M2-M5 system there is no stretched M2-brane. 
It is known that 
most questions in the typical irreducible representations reduce to 
those in the ordinary affine Lie algebra $\widehat{\mathfrak{g}}$. 
For example, it was shown in \cite{MR519631} 
that the classical Weyl-Kac character formula holds for 
arbitrary typical finite dimensional irreducible modules such that  
$\dim \widehat{\mathfrak{sg}}<\infty$ and $\dim L(\lambda)<\infty$. 
In the context of the AGT correspondence, 
the intersection of non-parallel M5-branes wrapping $\Sigma$ leads to 
a relation between instanton partition functions in the four-dimensional $\mathcal{N}=2$ quiver gauge theories 
in the presence of certain surface operators and 
conformal block of the affine Lie algebra $\widehat{\mathfrak{g}}$ 
\cite{Alday:2010vg, Kozcaz:2010yp, Negut:2011aa, Nawata:2014nca, Frenkel:2015rda}. 
Since the typical modules of $\widehat{\mathfrak{sg}}$ essentially contain 
the affine Lie algebra $\widehat{\mathfrak{g}}$,
and likewise the M2-M5 system realizes two intersecting M5-branes 
without any suspended M2-branes
as a special case (see Figure \ref{typ1}), 
it may be possible to extend the AGT correspondence, in the presence of surface operators as
a combination of M2-like and M5$'$-like surface operators, in terms of 
the affine Lie superalgebra $\widehat{\mathfrak{sg}}$. 
\noindent

A relation between brane configurations and atypical representations of a
supergroup has also been described previously by Mikhaylov and Witten
\cite{Mikhaylov:2014aoa}. In that case the supergroup arose from D3-branes
ending on both sides on an NS5-brane. The labels $\{x_i\}$ and $\{y_k\}$ were
associated with the D3-branes ending on the left and right, respectively, of
the NS5-brane. In the type IIB configuration dual to our M-brane construction,
we have the stacks of D3$^{+}$- and D3$^{-}$-branes on each side of an
NS5-brane. Since the M2-branes arise as a combination of these two stacks of
branes, it is consistent that the two sets of labels are both associated with
the same M2-branes. Also, as previously noted, the introduction of the M5-
and M5$'$-branes corresponds to NS5- and NS5$'$-branes which remove the freedom
for the D3-branes to move in the $34$ directions. Thus it should not be
surprising that in the limit we are considering the D3$^{+}$- and
D3$^{-}$-branes should have the same vacuum configuration, and hence the
$\{x_i\}$ and $\{y_k\}$ should be the same, giving maximum atypicality.

We could introduce further stacks of D3$^{+}$- and D3$^{-}$-branes on the other
side of the NS5$'$-brane. We would expect the case where some D3$^{+}$-branes
(and likewise for D3$^{-}$-branes)
on either side of the NS5$'$-brane carried the same charges to have special
properties. This would give the M-theory case where M2-branes ended on both
sides of the M5$'$-brane. However, further study of this is beyond the scope of this paper.

\section{Mock Modular Index}
\label{secmock}

\subsection{Definition}
\label{subseccurr}
We have identified the highest weight atypical module $L(\Lambda)$ over $\widehat{\mathfrak{gl}}(N|N)$ 
for given vacuum configuration of M2-M5 system. 
Now we want to study these modules via the indices and partition functions. 
We define an index for the supergroup WZW models by 
\begin{align}
\label{ind1a}
\mathcal{I}(\tau,z)&:=
\mathrm{Tr}_{\mathcal{H}}
(-1)^{F}q^{H_{L}}
\prod_{a=1}^{d}x_{a}^{F_{a}}.
\end{align}
Here $(-1)^{F}$ is the fermion number operator and 
$q:=e^{2\pi i\tau}$ is a complex parameter 
associated with the left-moving Hamiltonian $H_{L}=2(H+iP)=L_{0}-\frac{c}{24}$. 
The vector $F_{a}$ is the charge vector associated 
with the Cartan subalgebra for the atypical block 
of atypicality $d$ in the bosonic subalgebra $\mathfrak{g}_{\overline{0}}$, 
where $a=1,\cdots,d$. 
We have introduced the associated chemical potential $x_{a}:=e^{2\pi iz_{a}}$.
This index is an analogue of the Witten index for the supersymmetric quantum mechanics in that 
the $z_{a}\rightarrow 0$ limit gives the Witten index. 

Now we are ready to explain how the index (\ref{ind1a}) encodes the data of the M2-M5 system. 
We take the Hilbert space $\mathcal{H}$ as the irreducible atypical highest weight modules 
with atypicality $d$ being the number of the stretched M2-branes. 
The left-moving Hamiltonian $H_{L}$ is an energy of the sandwiched M2-branes,  
i.e. a winding number of the stretched M2-branes along 
one of the cycles of $\Sigma$, viewed as the Euclidean time circle.
The $F_{a}$, $a=1,\cdots,d$ are the $U(1)$ charges for a holomorphic $U(1)$ vector bundle 
over the Riemann surface wrapped by the stretched M2-branes, 
which originates from the 3-form $C$-field (\ref{c3field}). 
Therefore the index (\ref{ind1a}) counts BPS states of the M2-M5 system.

In addition, we consider a partition function 
\begin{align}
\label{ind1b}
\mathcal{Z}(\tau,\overline{\tau},z)&:=
\mathrm{Tr}_{\mathcal{H}}
(-1)^{F}q^{L_{0}-\frac{c}{24}}\overline{q}^{\overline{L}_{0}-\frac{c}{24}}
\prod_{a=1}^{d}x_{a}^{F_{a}}.
\end{align}
Here $\overline{q}^{\overline{L}_{0}-\frac{c}{24}}$ insert 
the right-moving Hamiltonian $H_{R}=2(H-iP)=\overline{L}_{0}-\frac{c}{24}$ 
into the index (\ref{ind1a}).
The partition function has the same form as the equivariant elliptic genus. 
It can be formulated by a path integral on a torus  
with a coordinate $w=\sigma_{1}+\tau\sigma_{2}$ 
where $\sigma_{1}$ and $\sigma_{2}$ are periodic with periodicity $2\pi$ and
$\tau$. Here $\tau=\tau_{1}+i\tau_{2}$ 
characterizes the complex structure of a torus $w\simeq w+2\pi\simeq w+2\pi\tau$, 
on which the WZW model is defined. 
From the point of view of the M2-M5 system, 
the right-moving Hamiltonian $H_{R}$ is a momentum of the stretched M2-branes 
along the other cycle of $\Sigma$, viewed as the Euclidean spatial circle.

A torus partition function should be the same for equivalent tori. 
A holomorphic function $\varphi$ on the upper half plane $\mathbb{H}$ transforming 
under the modular group $SL(2,\mathbb{Z})$ of reparametrizations 
of the torus as 
\begin{align}
\label{modular1}
\varphi
\left(\frac{a\tau+b}{c\tau+d}\right)&=
\left(c\tau+d\right)^{k}\varphi(\tau),& 
&
\left(
\begin{array}{cc}
a&b\\
c&d\\
\end{array}
\right)
\in SL(2,\mathbb{Z})
\end{align}
is called a modular form of weight $k$. 
The effect of a chemical potential $x_{a}$ is equivalent to 
the coupling of external gauge fields $A^{a}$ on the torus 
to the current so that 
\begin{align}
\label{ind3}
x_{a}&=e^{2\pi iz_{a}}=
e^{2\pi i
\left[\oint_{t}A^{a}-\tau\oint_{s}A^{a}\right]},
\end{align}
where $\oint_{t}$ (resp.\ $\oint_{s}$) is the temporal (resp.\ spatial) cycle of the torus. 
Such coupling is translated into the twisted boundary conditions 
of the fields $\phi(w,\overline{w})$ along the two cycles
\begin{align}
\label{ind3a}
\phi(w+2\pi,\overline{w}+2\pi)
&=\prod_{a}e^{2\pi i F_{a}\oint_{t}A^{a}}
\phi_{t}(w,\overline{w}),\\
\label{ind3b}
\phi(w+2\pi\tau,\overline{w}+2\pi\overline{\tau})
&=\prod_{a}e^{2\pi i\tau F_{a}\oint_{s}A^{a}}
\phi_{s}(w,\overline{w}),
\end{align}
where $\phi_{t}$ (resp.\ $\phi_{s}$) is the untwisted boundary condition 
along the temporal (resp.\ spatial) cycle. 
A function $\varphi(\tau,z)$ is called elliptic with index $m$ in $z$ 
if it has a transformation law 
\begin{align}
\label{ell1}
\varphi(\tau,z+\lambda\tau+\mu)&=
e^{-2\pi im(\lambda^{2}\tau+2\lambda z)}\varphi(\tau,z),& 
\lambda,\mu\in \mathbb{Z}
\end{align}
under the translation of $z$. 
A holomorphic function $\varphi(\tau,z)$ on $\mathbb{H}\times \mathbb{C}$ with the ellipticity (\ref{ell1}) 
which transforms under the modular group $SL(2,\mathbb{Z})$ as
\begin{align}
\label{jacob1}
\varphi
\left
(\frac{a\tau+b}{c\tau+d}, \frac{z}{c\tau+d}
\right)
&=
(c\tau+d)^{k} e^{\frac{2\pi imcz^{2}}{c\tau+d}}\varphi(\tau,d),& 
\left(
\begin{array}{cc}
a&b\\
c&d\\
\end{array}
\right)\in SL(2,\mathbb{Z})
\end{align}
is called a Jacobi form of weight $k$ and index $m$.

\subsection{Kac-Wakimoto formula}
\label{subsecaff}
In order to compute the indices, 
we recall the definition of the character $\mathrm{ch}_{L(\Lambda)}$ and 
the supercharacter $\mathrm{sch}_{L(\Lambda)}$ of the module $L(\Lambda)$ 
\begin{align}
\label{affch1}
\mathrm{ch}_{L(\Lambda)}&:=
\sum_{h \in \widehat{\mathfrak{h}}}
\mathrm{dim} L(\Lambda)
e^{h},& 
\mathrm{sch}_{L(\Lambda)}&:=
\sum_{h \in \widehat{\mathfrak{h}}}
\mathrm{sdim} L(\Lambda)
e^{h}.
\end{align}
The module $L(\Lambda)$ is integrable if and only if 
the character is invariant under 
$\widehat{W}^{\sharp}=W^{\sharp}\ltimes t_{L^{\sharp}}$, which is the subgroup of the affine Weyl group $\widehat{W}$ 
where $L^{\sharp}$ is the sublattice of the coroot lattice $L$ 
corresponding to the root system (\ref{rootsharp}).

Using the coordinate (\ref{affcar1b}) for $h\in \mathfrak{h}$, 
the supercharacter can be written explicitly as 
\begin{align}
\label{affsch1a0}
\mathrm{sch}_{L(\Lambda)}(\tau,z,t)
=\mathrm{Str}_{L(\Lambda)}
e^{2\pi i(-\tau d+z+tK)}.
\end{align}
It is demonstrated in \cite{MR1104219} that 
for an integrable $L(\Lambda)$ 
the supercharacter absolutely converges in the convex domain 
$D=\left\{
h\in \widehat{\mathfrak{h}}\ |\ 
\mathrm{Re}\ \alpha_{i}(\mathfrak{h})>0,i=1,\cdots,l
\right\}$
to a holomorphic function. 
Also, for all known examples 
it converges in the upper half plane 
$\mathbb{H}=\left\{
(\tau,z,t)\ |\ \mathrm{Im}\ \tau >0
\right\}$ 
to a meromorphic function.

Since the replacement of $\Lambda$ with 
$\Lambda+a\delta$ for $a\in \mathbb{C}$ keeps $L(\Lambda)$ irreducible, 
we further consider the supercharacter multiplied by $q^{a}$. 
The normalized supercharacter $\mathrm{sch}_{\Lambda}$ 
is defined by multiplying the supercharacter $\mathrm{sch}_{L(\Lambda)}$ by $q^{m_{\Lambda}}$ 
\cite{MR1104219} 
\begin{align}
\label{affnsch1a1}
\mathrm{sch}_{\Lambda}
&=q^{m_{\Lambda}}
\mathrm{sch}_{L(\Lambda)}(\tau,z,t)
\end{align}
where $m_{\Lambda}=
\frac{\left( \Lambda+\widehat{\rho},\Lambda+\widehat{\rho}\right)}
{2(k+h^{\vee})}
-\frac{\mathrm{sdim}\mathfrak{sg}}{24}=h_{\Delta}-\frac{c}{24}$ 
is called the modular anomaly. 
The normalized factor $q^{m_{\Lambda}}$ is necessary to realize the
contributions from the zero mode of the Virasoro generator $L_{0}$. 
It is associated to the modular invariance for the bosonic WZW models.
However, for the supergroup WZW models it is needed to acquire the intriguing
mock modular property, as we will see later.

%
The supercharacter formula for the atypical integrable module  
$L(\Lambda)$ given by the Kac-Wakimoto formula 
\cite{MR1327543,MR3392530} 
\begin{align}
\label{affsch1b1}
e^{\widehat{\rho}}
\widehat{R}^{-} \mathrm{sch}_{L(\Lambda)}
&=\sum_{w\in \widehat{W}^{\sharp}} \mathrm{sgn}^{-}(w) 
\frac{e^{w(\Lambda+\widehat{\rho})}}
{\prod_{\beta\in S}(1-e^{-w(\beta)})}
\end{align}
where 
\begin{align}
\label{affden1a}
\widehat{R}^{-}&=
\frac{\prod_{\alpha\in \widehat{\Delta}_{\overline{0}}^{+}}(1-e^{-\alpha})}
{\prod_{\alpha\in \widehat{\Delta}_{\overline{1}}^{+}}(1-e^{-\alpha})}
\end{align}
is the affine superdenominator, 
$\widehat{W}^{\sharp}=W^{\sharp}\ltimes t_{L^{\sharp}}$ is the subgroup of the affine Weyl group $\widehat{W}$,
$L^{\sharp}$ is the corresponding sublattice of the coroot lattice $L$,
and $\mathrm{sgn}^{-}(w)$ is the sign factor defined by (\ref{sign1}).

Furthermore from eqs.(\ref{affnsch1a1}) and (\ref{affsch1b1}) 
the normalized supercharacter is expressed as 
\cite{MR1810948,MR3200431}
\begin{align}
\label{affnsch1a2}
\mathrm{sch}_{\Lambda}&=
\frac
{\sum_{w\in W^{\sharp}}\mathrm{sgn}^{-}(w)  
\Theta_{\Lambda+\widehat{\rho},S}^{L^{\sharp},-}}
{q^{\frac{\mathrm{sdim}\mathfrak{sg}}{24}}\widehat{R}^{-}}. 
\end{align}
%
%
It turns out that 
the denominator in the formula (\ref{affnsch1a2}) consists of 
the theta functions $\vartheta_{11}$, $\vartheta_{10}$ and powers of the
eta function $\eta(\tau)$ (see Appendix \ref{appjacobi}), 
which are members of a modular invariant family. 
On the other hand, the function $\Theta_{\Lambda, S}^{Q,\pm}$ in the numerator 
is a Ramanujan mock theta function 
\cite{MR2280843, MR1567721, MR1542896} 
defined as the series 
\cite{MR3200431,MR3534829,MR3535359}
\begin{align}
\label{affmock1a}
\Theta_{\Lambda, S}^{Q,\pm}
&=q^{-\frac{2(\Lambda,\Lambda)}{2K}\delta}
\sum_{\gamma\in Q}
\mathrm{sgn}^{\pm}(t_{\gamma})
\frac{e^{t_{\gamma}(\Lambda)}}
{\prod_{\beta\in S}(1-e^{-t_{\gamma}(\beta)})}
\end{align}
where $t_{\gamma}$ is the element of the affine Weyl group $\widehat{W}$ defined in (\ref{aweyl2}). 
The mock theta function $\Theta^{Q,\pm}_{\Lambda,S}$ is determined by four data; 
(i) the weight $\Lambda\in \widehat{\mathfrak{h}}^{*}$ with $\Lambda(K)>0$, 
(ii) the positive definite integral root lattice $Q$ of $\mathfrak{h}^{*}_{\mathbb{R}}$,  
(iii) the finite subset $S \subset \widehat{\mathfrak{h}}_{\mathbb{R}}^{*}$ 
composed of pairwise orthogonal isotropic vectors orthogonal to $\Lambda$, and 
(iv) the homomorphism $\mathrm{sgn}^{\pm}(\gamma):Q\rightarrow \{\pm\}$, with $\gamma\in Q$. 
The degree of the mock theta function (\ref{affmock1a}) is $\Lambda(K)=k$ 
and the $\Theta_{\Lambda+\widehat{\rho}, S}^{L^{\sharp},-}$ 
in the Kac-Wakimoto formula (\ref{affnsch1a2}) is a mock theta function 
of degree $k+h^{\vee}$. 
%
%
%

\subsection{Computation}
\label{subsecaffComp}
%
Comparing (\ref{ind1a}) with (\ref{affsch1a0}), 
we find that the index (\ref{ind1a}) is the specialization 
of the supercharacter
\begin{align}
\label{0index1a10}
\mathcal{I}(\tau,z)&=
\mathrm{sch}_{\Lambda}(\tau,z,0)
\end{align}
for $k=1$. 
From now on we restrict our attention to the atypical module $L(\Lambda)$ 
and take it as the Hilbert space $\mathcal{H}$ in the definition of the indices. 
Applying the Kac-Wakimoto formula (\ref{affnsch1a2}), 
we see that the index $\mathcal{I}(\tau,z)$ can be expressed in terms of the mock theta function. 
We thus call this index, which is analogous to the Witten index, a mock modular index.

Next, consider the torus partition function 
$\mathcal{Z}(\tau,\overline{\tau},z)$. 
For the equivariant elliptic genus in compact superconformal field theories,
the Hilbert spaces only contain discrete sets of primary fields.
The additional factor $\overline{q}^{\overline{L}_{0}-\frac{c}{24}}$
requires the combined left and right moving sectors.
However, there is a cancellation between bosonic and fermionic fluctuations 
from supersymmetry. Then, due to the discreteness of the spectrum in the Ramond sector, 
there is just an algebraic sum of the spectrum in the Ramond sector,
and the contribution only arises from the ground states of the Ramond sector. 
This ensures the holomorphicity of the elliptic genus. 

However, the emergence of the mock theta function does not allow us to 
extend $\mathcal{I}(\tau,z)$ to $\mathcal{Z}(\tau,\overline{\tau},z)$ 
by naively inserting the factor $\overline{q}^{\overline{L}_{0}-\frac{c}{24}}$ 
without any modification of the result. 
This is because the index $\mathcal{Z}(\tau,\overline{\tau},z)$ should be modular invariant 
due to the path integral formalism while the index $\mathcal{I}(\tau,z)$ is not. 
This indicates that some pieces in $\mathcal{Z}(\tau,\overline{\tau},z)$, are missing in $\mathcal{I}(\tau,z)$ 
and a proper completion must be added to restore the modular invariant $\mathcal{Z}(\tau,\overline{\tau},z)$.

Such a property of the spectrum stems from the structure of the Hilbert space $\mathcal{H}$ of the theory under consideration. 
The holomorphic elliptic genus relies on the fact that 
$\mathcal{H}$ has a holomorphically factorized form 
\begin{align}
\label{hilbert1a}
\mathcal{H}&=
\bigoplus_{\mu} 
\mathcal{H}_{\mu}\otimes \overline{\mathcal{H}}_{\mu}
\end{align}
where $\mathcal{H}_{\mu}$ (resp.\ $\overline{\mathcal{H}}_{\mu}$) is 
the holomorphic (resp.\ anti-holomorphic) sector. 
However, for the supergroup WZW models 
the space of the states has been argued to have the form 
\cite{Schomerus:2005bf,Quella:2007hr,Mitev:2011zza,Quella:2013oda}
\begin{align}
\label{hilbert1b}
\mathcal{H}&=
\left(\ 
\bigoplus_{\mu\in \textrm{typical}}
\mathcal{H}_{\mu}\otimes \overline{\mathcal{H}_{\mu}}
\ \right)
 \oplus 
 \left(\ 
\bigoplus_{\nu \in\textrm{atypical}}
\widehat{\mathcal{H}}_{\nu}
\ \right).
\end{align}
Although there is the holomorphic factorization 
$\mathcal{H}_{\mu}\otimes \overline{\mathcal{H}_{\mu}}$
in the typical sector, 
in the atypical sector $\widehat{\mathcal{H}}_{\nu}$, 
the holomorphic and anti-holomorphic parts 
are entangled with each other in a complicated way. 
This observation is consistent with our conclusion 
as we are now dealing with $\widehat{\mathcal{H}}_{\nu}$, 
the Hilbert space of an atypical module.

The appearance of the mock theta function $\Theta_{\Lambda, S}^{Q,\pm}$ 
in the normalized supercharacter is remarkable in that 
although the mock modular functions are not exactly modular invariant, 
they can be made modular invariant 
by adding suitable non-holomorphic completions  
developed by Zwegers \cite{Zwegers:2008zna}. 
The basic idea is that 
a new non-holomorphic function 
\begin{align}
\label{zwegers1a}
\widehat{h}(\tau,\overline{\tau})&=
h(\tau)+{g}^{*}(\tau,\overline{\tau}),
\end{align}
created by the addition of the non-holomorphic Eichler integral 
\begin{align}
\label{zwegers2}
g^{*}&=\left(
\frac{i}{2\pi}
\right)^{k-1}\int_{-\overline{\tau}}^{\infty}dz
(z+\tau)^{-k}\overline{g(-\overline{z})}
\end{align}
constructed from a holomorphic modular form $g(\tau)$ of weight $2-k$, called a shadow of $h(\tau)$, 
turns out to be modular invariant at the cost of holomorphicity. 
This naturally leads to a prescription for the evaluation of 
the non-holomorphic part of the modular invariant partition function $\mathcal{Z}(\tau,\tau,z)$ 
defined by (\ref{ind1b}) on an elliptic curve as
\begin{align}
\label{0index1a1}
\mathcal{Z}(\tau,\overline{\tau},z)&=\widehat{\mathcal{I}}_(\tau,\overline{\tau},z)
+(\textrm{holomorphic modular function}).
\end{align}
The first term $\widehat{\mathcal{I}}(\tau,\overline{\tau},z)$ 
is the modular completion of $\mathcal{I}(\tau,z)$ via 
Zwegers' method (\ref{zwegers1a}), which is the contribution 
from the atypical sector $\widehat{\mathcal{H}}_{\nu}$,
while the remnant is the holomorphic modular function arising from 
the typical sector $\mathcal{H}_{\mu}\otimes \overline{\mathcal{H}}_{\mu}$. 
Note that the index $\mathcal{Z}(\tau,\overline{\tau},z)$ is no longer holomorphic 
due to $\widehat{\mathcal{I}}(\tau,\overline{\tau},z)$ but it is modular invariant.

\subsection{$PSL(2|2)_{k=1}$ WZW model}
\label{subsecpsl22}
\begin{figure}[H]
\begin{center}
\includegraphics[width=4.5cm]{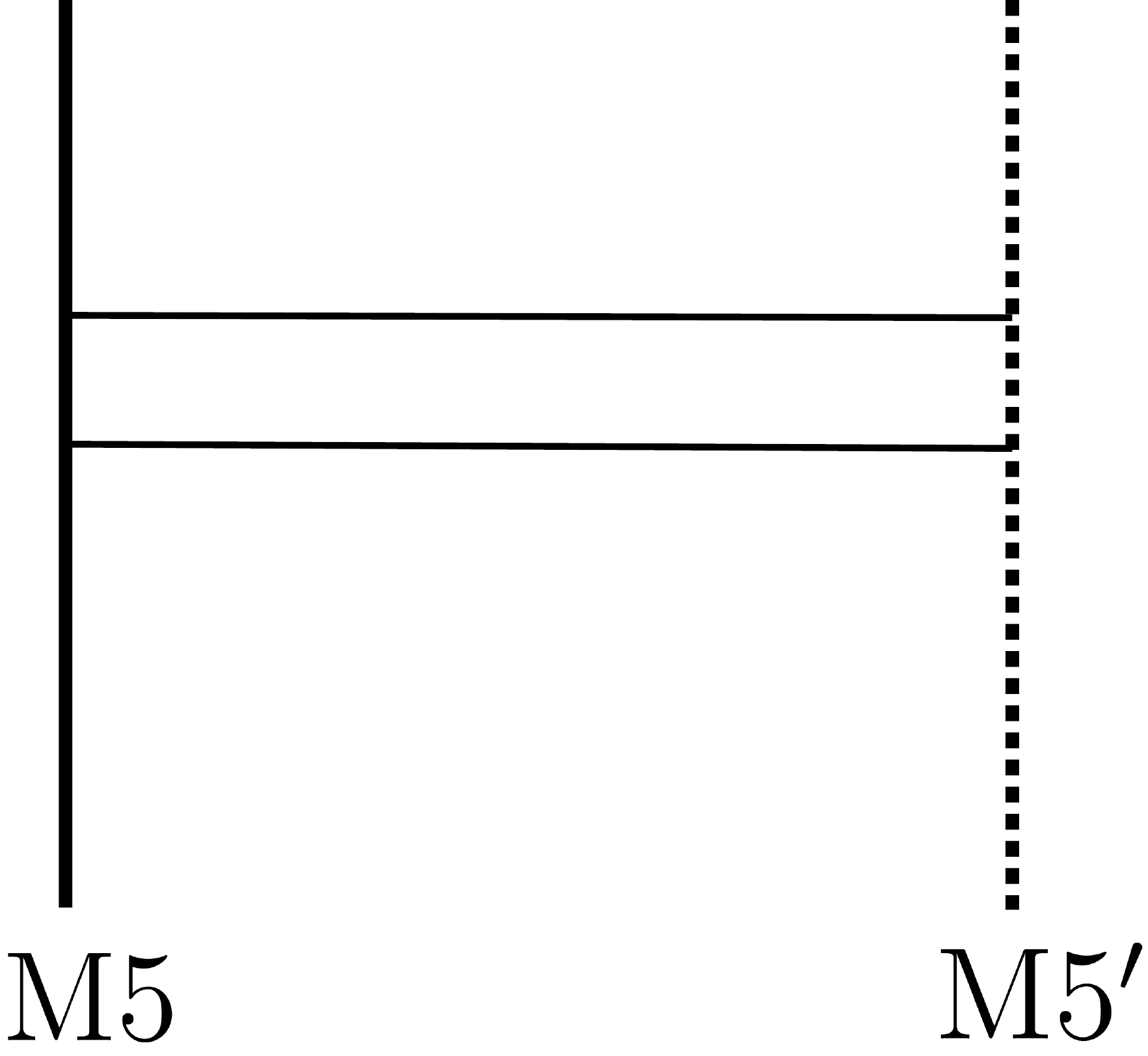}
\caption{$N=d=2$ M2-branes stretched between the M5- and M5$'$-branes. }
\label{mdiagram2}
\end{center}
\end{figure}

In this subsection 
we will provide a simple example of the index computation 
for the $PSL(2|2)_{k=1}$ WZW model. 
The corresponding brane configuration is illustrated in Figure \ref{mdiagram2} 
where $N=d=2$ M2-branes are stretched between the M5- and M5$'$-brane. 
\noindent
For example, given M2-brane charges 
$\{x_{i}\}=\{4,2\}$ and 
$\{y_{k}\}=\{2,4\}$, 
the weight of the irreducible highest weight module with maximal atypicality $d=2$ is given by
\begin{align}
\label{bdia1a0}
\lambda+\rho&=
4\epsilon_{1}+2\epsilon_{2}
-2\delta_{1}-4\delta_{2}
\end{align}
and the weight diagram has only $\vee$'s and $\wedge$'s as follows:
\begin{align}
\label{bdia1a1}
\begin{array}{ccccccc}
\wedge&\wedge&\vee&\wedge&\vee&\wedge&\wedge\\
0&1&2&3&4&5&6 \\
\end{array}.
\end{align}

\begin{figure}
\centering
    \begin{tabular}{cc}
          \includegraphics[scale=0.25]{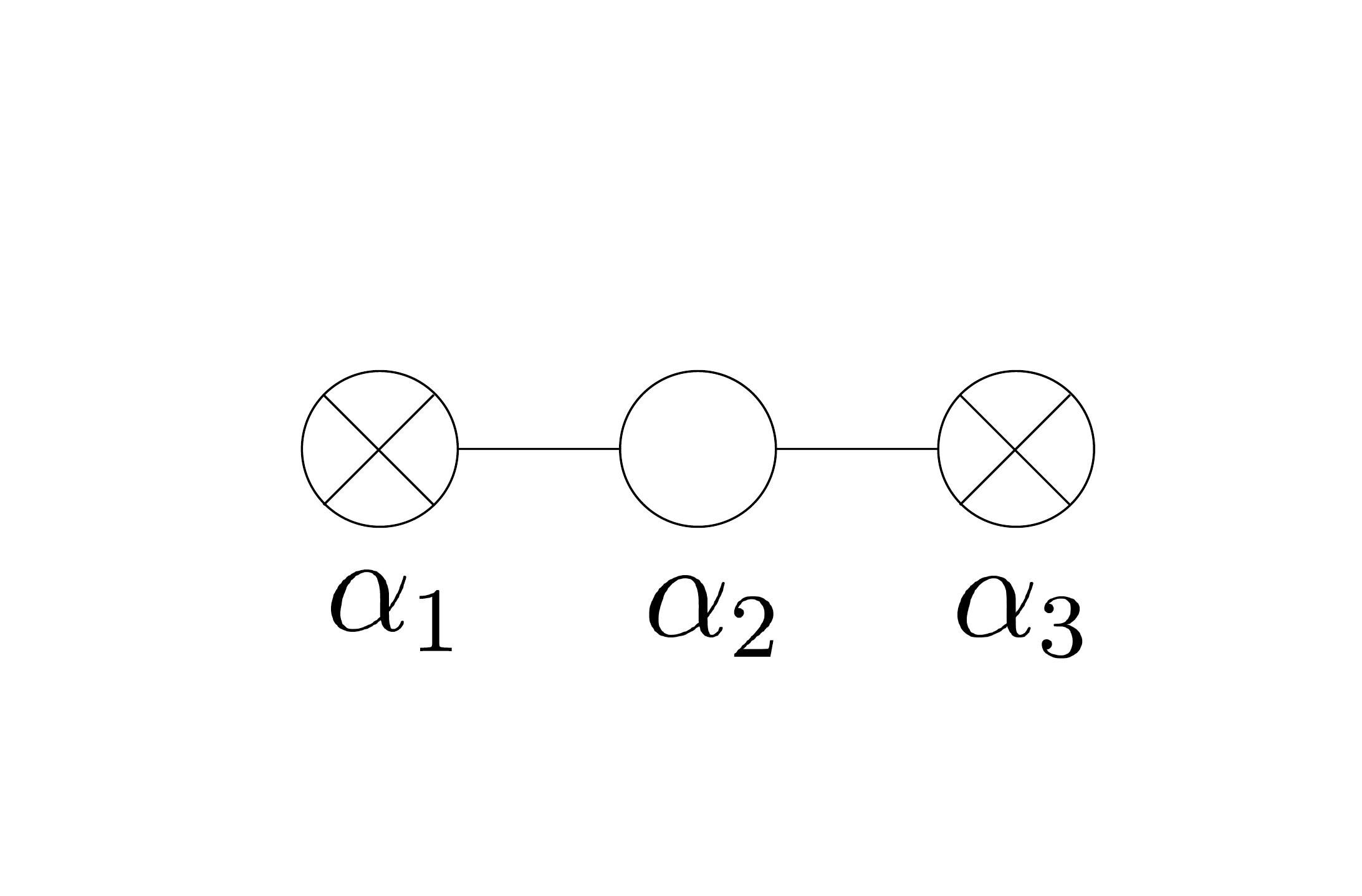}  & \includegraphics[scale=0.25]{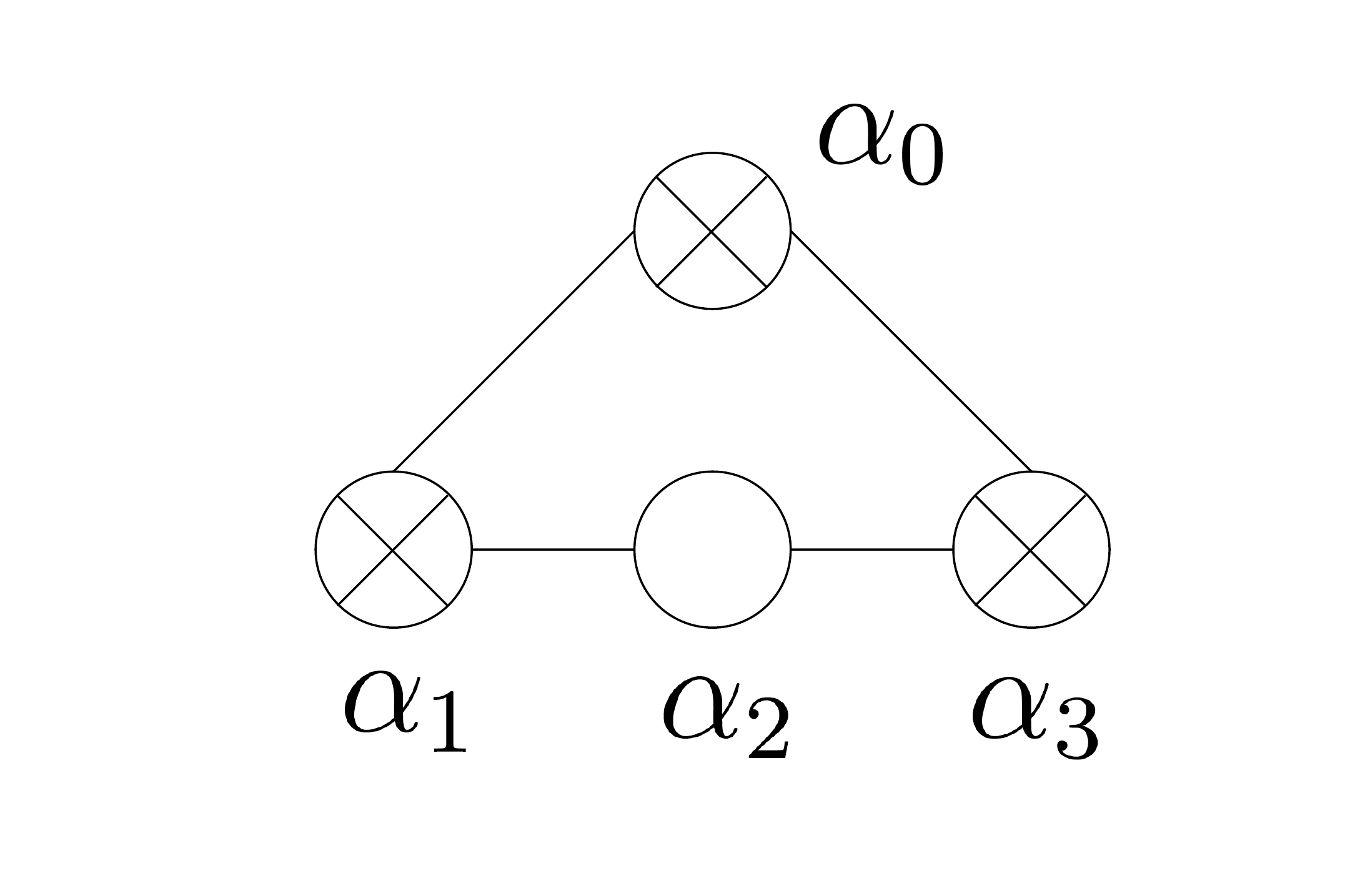}\\
          (a) $\mathfrak{psl}(2|2)$. &(b) $\widehat{\mathfrak{psl}}(2|2)$. 
\end{tabular}
    \caption{The Dynkin diagrams corresponding to (\ref{psl22simple}). 
    The white dot $\bigcirc$ represents a simple even root and the gray dot $\bigotimes$ represents a simple odd root of zero length. }
    \label{dyn1}
\end{figure}

The Cartan subalgebra $\widehat{\mathfrak{h}}$ 
of $\widehat{\mathfrak{psl}}(2|2)$ 
takes the form of (\ref{affcar1a}) where 
$\mathfrak{h}$ is the quotient of 
diagonal matrices of $\mathfrak{sl}(2|2)$ by $\mathbb{C}I_{4}$. 
We choose a simple root system of $\mathfrak{psl}(2|2)$ as
\begin{align}
\label{psl22simple}
\Pi&=\{\alpha_{1},\alpha_{2},\alpha_{3}\}=\{\epsilon_{1}-\delta_{1}, \delta_{1}-\delta_{2}, \delta_{2}-\epsilon_{2}\}
\end{align}
where $\alpha_{1}=\alpha_{3}$. 
The corresponding Cartan matrix is 
\begin{align}
\label{psl22cartan}
\left(
\begin{array}{ccc}
0&1&0\\
1&-2&1\\
0&1&0\\
\end{array}
\right)
\end{align}
and the Dynkin diagram is shown in Figure \ref{dyn1}. 
We then have inner products 
\begin{align}
\label{2root1a}
(\alpha_{1},\alpha_{1})&=( \alpha_{3},\alpha_{3})=( \alpha_{1},\alpha_{3})=0,& 
( \alpha_{2},\alpha_{2})&=-2,\nonumber\\
( \alpha_{1},\alpha_{2})&=
( \alpha_{2},\alpha_{3})=1,& ( \theta,\theta)&=2
\end{align}
where $\theta=\alpha_{1}+\alpha_{2}+\alpha_{3}=\epsilon_{1}-\epsilon_{2}$ is a highest root. 
The positive root systems and the Weyl vectors of $\mathfrak{psl}(2|2)$ are
\begin{align}
\label{2root1b}
\Delta_{\overline{0}}^{+}&=
\{\alpha_{2},\theta\}=\{\delta_{1}-\delta_{2}, \epsilon_{1}-\epsilon_{2} \},\\
\label{2root1b1}
\Delta_{\overline{1}}^{+}&=
\{\alpha_{1},\alpha_{3},\alpha_{12},\alpha_{23}\}=
\{\epsilon_{1}-\delta_{1}, \delta_{2}-\epsilon_{2}, \epsilon_{1}-\delta_{2}, \delta_{1}-\epsilon_{2}\},\\
\label{2root1c}
\rho_{\overline{0}}&=\frac12 (\alpha_{12}+\alpha_{23})
=\frac12 (\epsilon_{1}-\epsilon_{2}+\delta_{1}-\delta_{2}),
\\
\label{2root1c1}
\rho_{\overline{1}}&=\theta=\alpha_{123}=\epsilon_{1}-\epsilon_{2},
\\
\label{2root1c2}
\rho&= \rho_{\overline{0}} - \rho_{\overline{1}} = -\frac12 \alpha_{13}
= -\frac12 (\epsilon_{1}-\epsilon_{2}-\delta_{1}+\delta_{2})
\end{align}
where $\alpha_{ij}:=\alpha_{i}+\alpha_{j}$ and $\alpha_{ijk}:=\alpha_{i}+\alpha_{j}+\alpha_{k}$. 
Let us choose a coordinate (\ref{affcar1b}) on $\widehat{\mathfrak{h}}$ as
\begin{align}
\label{2affcar1b}
h&:=
2\pi i\left(-\tau d-(z_{1}+z_{2})\alpha_{1}-z_{1}\alpha_{2}+tK\right)
\end{align}
where $\tau$, $z_{1}$, $z_{2}$, $t\in \mathbb{C}$ 
and $z:=-(z_{1}+z_{2})\alpha_{1}-z_{1}\alpha_{2}$ is a coordinate on $\mathfrak{h}$ 
with an inner product $(z,z)=2z_{1}z_{2}$. 
. 
%

%
%
%
%
%
%


For $\widehat{\mathfrak{psl}}(2|2)$ 
the normalized affine superdenominator (\ref{affden1a}) 
is expressed as \cite{MR3200431}
\begin{align}
\label{2affden1a}
\widehat{R}^{-}(\tau,z_{1},z_{2})
&=\eta(\tau)^{4}
\frac{\vartheta_{11}(\tau,z_{1}-z_{2})\vartheta_{11}(\tau,z_{1}+z_{2})}
{\vartheta_{11}(\tau,z_{1})^{2}\vartheta_{11}(\tau,z_{2})^{2}}
\end{align}
where $\eta(\tau)$ is the Dedekind eta function (\ref{eta1a}) 
and $\vartheta_{11}(\tau,z)$ is the Jacobi theta function (\ref{theta1d}). 
Since $\mathfrak{psl}(2|2)$ has zero dual Coxeter number,　
$\widehat{R}^{-}(\tau,z_{1},z_{2})$ has 
no dependence on parameter $t\in \mathbb{C}$ in (\ref{2affcar1b}). 
From eqs.(\ref{eta1b}), (\ref{theta2c}) and (\ref{theta3c}), 
the modular transformations of 
$\widehat{R}^{-}(\tau,z_{1},z_{2})$ read
\begin{align}
\label{2affden1a1}
\widehat{R}^{-}
\left(-\frac{1}{\tau},\frac{z_{1}}{\tau},\frac{z_{2}}{\tau}\right)
&=i\tau e^{\frac{\pi iz_{1}z_{2}}{\tau}}\widehat{R}^{-}(\tau,z_{1},z_{2}),\\
\label{2affden1a2}
\widehat{R}^{-}\left(\tau+1,z_{1},z_{2}\right)
&=e^{-\frac{\pi i}{3}}\widehat{R}^{-}(\tau,z_{1},z_{2}).
\end{align}
%
Following \cite{MR3200431}, we here consider the normalized supercharacter of 
the atypical module $L(\Lambda)$ for $\Lambda$ admissible
\cite{MR949675, MR1026952}\footnote{
It has been conjectured in \cite{MR949675, MR1026952} that 
if the highest weight module $L(\Lambda)$ is modular invariant, 
$\Lambda$ is realized as an admissible weight. }.  
The admissible weight $\Lambda$ is classified 
by the so-called simple subset $S={\varphi^{*}}^{-1}(\widehat{\Pi})$ 
$\in \widehat{\Delta}_{+}$ \cite{MR1026952} 
for the compatible homomorphism $\varphi: \widehat{\mathfrak{sg}}\rightarrow \widehat{\mathfrak{sg}}$. 
Let $\varphi(K)=MK$ where $M$ is a positive integer called the degree of $\varphi$. 
For $\widehat{\mathfrak{psl}}(2|2)$ 
the conditions for the admissible weights are given by \cite{MR3200431}
\begin{align}
\label{2afflevel1}
K&=\frac{m}{M},&
\mathrm{gcd}(m,M)&=1
\end{align}
where $m$ is a non-zero integer.  
There exist four simple subsets $S$ \cite{MR3200431}
\begin{align}
\label{2affset1}
S_{1}=
\{&
k_{i}\delta+\alpha_{i} 
|i=0,\cdots,3,\ \sum_{i=0}^{3}k_{i}=M-1 \}
,\nonumber\\
S_{2}=
\{&
k_{i}\delta-\alpha_{i} 
|i=0,\cdots,3,\ \sum_{i=0}^{3}k_{i}=M+1,\ k_{i}>0 \}
,\nonumber\\
S_{3}=\{&
k_{0}\delta+\alpha_{0}, 
k_{1}\delta+\alpha_{12}, 
k_{2}\delta-\alpha_{2}, 
k_{3}\delta+\alpha_{23}
|\sum_{i=0}^{3}k_{i}=M-1,k_{2}>0 \}
,\nonumber\\
S_{4}=\{&
k_{0}\delta-\alpha_{0}, 
k_{1}\delta-\alpha_{12}, 
k_{2}\delta+\alpha_{2}, 
k_{3}\delta-\alpha_{23}
|\sum_{i=0}^{3}k_{i}=M-1,k_{2}>0 \},
\end{align}
where we have introduced the integers  
$k_{i}\in \mathbb{Z}_{\ge0}$, $i=0,1,2,3$ with $k_{1}=k_{3}$. 
Setting $(j,k):=(k_{1}, k_{1}+k_{2})$, $j,k \in \mathbb{Z}_{\ge0}$, 
we obtain all the possible admissible highest weights $\Lambda_{jk}$ labelled by $(j,k)$ as follows: 
\begin{align}
\label{2affweight1}
\Lambda_{jk}&=
\begin{cases}
k\ge j\ge 0,\ \ j+k\le M-1&\textrm{for $s=1$}\cr
M-1\ge j\ge k\ge1,\ \ j+k\ge M&\textrm{for $s=2$}\cr
0\le k<j,\ \ j+k\le M-1&\textrm{for $s=3$}\cr
1\le j\le k\le M-1,\ \ j+k\ge M&\textrm{for $s=4$}.\cr
\end{cases}
\end{align}
with $s=1,2,3,4$ labelling the four simple subsets (\ref{2affset1}). 
Collecting all the results, 
the Kac-Wakimoto supercharacter formula (\ref{affnsch1a2}) 
for the admissible representations $\Lambda_{jk}$ of $\widehat{\mathfrak{psl}}(2|2)$ reads 
\cite{MR3200431} 
\begin{align}
\label{2affnsch1}
\mathrm{sch}_{\Lambda_{jk}}&=
\frac{
(-1)^{\frac{(s-1)(s-2)}{2}}
q^{\frac{mjk}{M}}e^{\frac{2\pi im}{M}}
\Phi^{[m]}
\left( M\tau, z_{1}+j\tau, z_{2}+k\tau, \frac{\tau}{M} \right)
}
{\widehat{R}^{-}}
\end{align}
where 
\begin{align}
\label{2affnsch2}
\Phi^{[m]}(\tau,z_{1},z_{2},t)&=
e^{2\pi nmt}\sum_{n\in \mathbb{Z}}
\left[
\frac{e^{2\pi inm(z_{1}+z_{2})}e^{2\pi iz_{1}}q^{mn^{2}+n}}
{(1-e^{2\pi iz_{1}}q^{n})^{2}}
-
\frac{e^{-2\pi inm(z_{1}+z_{2})}e^{-2\pi iz_{2}}q^{mn^{2}+n}}
{(1-e^{-2\pi iz_{2}}q^{n})^{2}}
\right]. 
\end{align}
To proceed with the index computation of the $PSL(2|2)_{k=1}$ WZW models, 
we first observe that 
the fixed level $k=1$ requires that the degree $M$ is equal to one. 
Furthermore the conditions (\ref{2afflevel1}), (\ref{2affset1}) and (\ref{2affweight1}) 
are realized only when $K=M=m=1$, $(j,k)=(0,0)$ for $s=1$. 
Making use of the formulae (\ref{0index1a10}) and (\ref{2affnsch1}), 
we obtain the mock modular index $\mathcal{I}(\tau,z)$ 
for the $PSL(2|2)_{k=1}$ WZW model
\begin{align}
\label{2affnsch1a}
\mathcal{I}(\tau,z_{1},z_{2})
&=
\frac{1}{\eta(\tau)^{4}}
\frac{\vartheta_{11}(\tau,z_{1})^{2}\vartheta_{11}(\tau,z_{2})^{2}}
{\vartheta_{11}(\tau,z_{1}-z_{2})\vartheta_{11}(\tau,z_{1}+z_{2})}\nonumber\\
&
\times \sum_{n\in \mathbb{Z}}
\left[
\frac{e^{2\pi in(z_{1}+z_{2})}e^{2\pi iz_{1}}q^{n^{2}+n}}
{(1-e^{2\pi iz_{1}}q^{n})^{2}}
-
\frac{e^{-2\pi in(z_{1}+z_{2})}e^{-2\pi iz_{2}}q^{n^{2}+n}}
{(1-e^{-2\pi iz_{2}}q^{n})^{2}}
\right].
\end{align}

\section{Appell-Lerch Sums}
\label{secALsums}
The holomorphic index (\ref{2affnsch1a}) takes the form:
\begin{align}
\label{appindex1}
\mathcal{I}&=
\frac{1}{\eta^4(\tau)}\frac
{
\vartheta_{11}^2(z_1; \tau)
\vartheta_{11}^2(z_2; \tau)
}
{\vartheta_{11}(z_1-z_2; \tau)
\vartheta_{11}(z_1+z_2; \tau)} \left( \mathcal{A}_{2,1}(\tau, z_1, z_1+z_2) 
-\mathcal{A}_{2,1}(\tau, -z_2, -z_1-z_2) \right)
\end{align}
where the second order Appell-Lerch sum is given by
\begin{align}
\label{appindex2}
\mathcal{A}_{2,1}(\tau, u, v) = U \sum_{n \in \mathbb{Z}} \frac{q^{n(n+1)}V^n}{(1 - Uq^n)^2}
\end{align}
and we have denoted $U = \exp(2\pi i u)$ and $V = \exp(2\pi i v)$. As
previously noted for the atypical modules, the issue, which we will now address,
is that the Appell-Lerch sums are not modular.

Following closely the method in \cite{Ashok:2014nua}, based on
\cite{Zwegers:2008zna, Dabholkar:2012nd}, 
we can complete the second order Appell-Lerch sums. 
The idea is to expressか the second order sum 
as a derivative of a first order sum. 
It is already known how to complete the first order sum,
so replacing it by its modular completion 
gives the modular completion of the second order sum, 
once we have taken into account the modular transformation properties 
coming from the derivative operator.

Explicitly, the modular completion of the first order Appell-Lerch sums
\begin{align}
\label{app1a}
\mathcal{A}_{1,k}(\tau, u, v)&
=U^k \sum_{n \in \mathbb{Z}} \frac{q^{kn(n+1)}V^n}{1 - Uq^n}
\end{align}
are the weight $1$ Jacobi forms
\begin{align}
\label{app1b}
\hat{\mathcal{A}}_{1,k}(\tau, u, v)&=
\mathcal{A}_{1,k}(\tau, u, v) + \mathcal{R}_{1,k}(\tau, u, v)
\end{align}
where
\begin{align}
\label{app1c}
\mathcal{R}_{1,k}(\tau, u, v)&
=\frac{i}{4k}U^{k - 1/2} \sum_{m = 0}^{2k-1} 
\vartheta_{11}\left( \frac{v+m}{2k} + \frac{(2k-1)\tau}{4k} ; \frac{\tau}{2k} \right) \nonumber \\
 &\times R\left( u - \frac{v+m}{2k} - \frac{(2k-1)\tau}{4k} ;
 \frac{\tau}{2k} \right) \\
\label{app1d}
R(w ; \tau)&=
\sum_{\nu \in \mathbb{Z} + 1/2} 
\left( \mathrm{sgn}(\nu) - \mathrm{Erf}
\left( \sqrt{2\pi\tau_2}\left( \nu + \frac{\Im(w)}{\tau_2} \right) \right) \right) \nonumber \\
 &\times (-1)^{\nu-1/2}W^{-\nu}q^{-\nu^2/2}
\end{align}
and $\tau_2 = \Im(\tau)$.

Now it is simple to check that
\begin{align}
\label{app1e}
\mathcal{D} \mathcal{A}_{1,k}(\tau, u, v)&=
(k-1)\mathcal{A}_{1,k}(\tau, u, v) + U^k \sum_{n \in \mathbb{Z}} \frac{q^{kn(n+1)}V^n}{(1 - Uq^n)^2}
\end{align}
where we define
\begin{align}
\label{app1f}
\mathcal{D}&=\frac{1}{2 \pi i}\frac{\partial}{\partial u}. 
\end{align}
So, we have for $k=1$ the simple relation
\begin{align}
\label{app1g}
\mathcal{A}_{2,1}(\tau, u, v)&
=\mathcal{D} \mathcal{A}_{1,1}(\tau, u, v). 
\end{align}
Since the modular transform of $\hat{\mathcal{A}}_{1,k}$ is
\begin{align}
\label{app1h}
\hat{\mathcal{A}}_{1,k}\left( \frac{a\tau+b}{c\tau+d},
 \frac{u}{c\tau+d}, \frac{v}{c\tau+d} \right)&
=(c\tau+d) \exp\left( \frac{2\pi ic}{c\tau+d} u(v-ku) \right) 
\hat{\mathcal{A}}_{1,k}(\tau, u, v), 
\end{align}
we can easily see that there is an extra term in the transformation of the
derivative. 
Specifically,
\begin{align}
\label{app1i}
\mathcal{D} \hat{\mathcal{A}}_{1,k}\left( \frac{a\tau+b}{c\tau+d},
 \frac{u}{c\tau+d}, \frac{v}{c\tau+d} \right) 
&=(c\tau+d) \exp\left( \frac{2\pi ic}{c\tau+d} u(v-ku) \right) \mathcal{D} \hat{\mathcal{A}}_{1,k}(\tau, u, v) \nonumber \\
&+c(v-2ku)\exp\left( \frac{2\pi ic}{c\tau+d} u(v-ku) \right) \hat{\mathcal{A}}_{1,k}(\tau, u, v)
\end{align}
but then it is easy to see that by shifting the derivative operator we get
the following expression which transforms as a weight $2$ Jacobi form:
\begin{align}
\left( \mathcal{D} + \frac{\Im(v)}{\tau_2} -2k\frac{\Im(u)}{\tau_2}
 \right) \hat{\mathcal{A}}_{1,k}(\tau, u, v) .
\end{align}
Combining the above results we see that the modular completion of
$\mathcal{A}_{2,1}(\tau, u, v)$ is
\begin{align}
\hat{\mathcal{A}}_{2,1}(\tau, u, v)&
=\left( \mathcal{D} + \frac{\Im(v)}{\tau_2} -2\frac{\Im(u)}{\tau_2} \right) \hat{\mathcal{A}}_{1,1}(\tau, u, v) .
\end{align}

Note that this works for the index since for both cases $u=z_1$, $v=z_1+z_2$ and
$u=-z_2$, $v=-z_1-z_2$ we see that $u(v-u) = z_1z_2$. So, the combination
$$ \hat{\mathcal{A}}_{2,1}(\tau, z_1, z_1+z_2) - \hat{\mathcal{A}}_{2,1}(\tau, -z_2, -z_1-z_2) $$
also transforms as a Jacobi form of weight $2$ (with index $1$), i.e.\ with a
factor
$$ (c\tau+d)^2 \exp\left( \frac{2\pi ic}{c\tau+d} z_1z_2 \right) $$
under a modular transformation.

If we include the $\vartheta$ and $\eta$ factors the whole completed index
\begin{align}
\hat{\mathcal{I}}&
=\frac{1}{\eta^4(\tau)}\frac{\vartheta_{11}^2(z_1; \tau)\vartheta_{11}^2(z_2; \tau)}
{\vartheta_{11}(z_1-z_2; \tau)\vartheta_{11}(z_1+z_2; \tau)} \left( \hat{\mathcal{A}}_{2,1}(\tau, z_1, z_1+z_2) - \hat{\mathcal{A}}_{2,1}(\tau, -z_2, -z_1-z_2) \right)
\label{ModIndex}
\end{align}
transforms as a Jacobi form of weight $1$ and index $1$.

Now to analyse the result, we define
\begin{align}
\label{app21a}
\mathcal{R}_{2,1}(\tau, u, v)
&=\hat{\mathcal{A}}_{2,1}(\tau, u, v) - \mathcal{A}_{2,1}(\tau, u, v) 
\nonumber\\
&=\mathcal{D}\mathcal{R}_{1,1}(\tau, u, v)+\left( \frac{\Im(v)}{\tau_2} - 2\frac{\Im(u)}{\tau_2} \right) \hat{\mathcal{A}}_{1,1}(\tau, u, v).
\end{align}

\subsection{Holomorphic anomaly}
\label{subsechol}
The completed index is not holomorphic and we can calculate a holomorphic
anomaly equation by taking its $\overline{\tau}$ derivative. Specifically, we can
calculate:
\begin{align}
\frac{\partial}{\partial \overline{\tau}} \hat{\mathcal{A}}_{2,1}(\tau, u, v)
 &=\frac{\partial}{\partial \overline{\tau}} \mathcal{R}_{2,1}(\tau, u,
 v)
\nonumber \\
 &=\frac{\partial}{\partial \overline{\tau}} \left(
 \mathcal{D}\mathcal{R}_{1,1} \right) + \left( \Im(v) - 2 \Im(u) \right)
 \frac{\partial}{\partial \overline{\tau}} \left( \frac{1}{\tau_2}
 \hat{\mathcal{A}}_{1,1}(\tau, u, v) \right) 
\nonumber \\
 &=\frac{-i}{2\tau_2^2} \left( \Im(v) - 2 \Im(u) \right)
 \hat{\mathcal{A}}_{1,1}(\tau, u, v) 
+ \left( \mathcal{D} + \frac{\Im(v) - 2 \Im(u)}{\tau_2} \right) \frac{\partial}{\partial \overline{\tau}} \mathcal{R}_{1,1} .
\label{nonholA21}
\end{align}
From the definition of $\mathcal{R}_{1,1}$ and noting that
\begin{align}
\label{erf1a}
\frac{d}{dz}\mathrm{Erf}(z) = \frac{2}{\sqrt{\pi}}e^{-z^2}
\end{align}
we find
\begin{align}
\label{hol1a}
\frac{\partial}{\partial \overline{\tau}} \mathcal{R}_{1,1}(\tau, u, v)
 & =\frac{1}{8\sqrt{2\tau_2}} e^{\pi i u} \sum_{m=0}^1 \vartheta_{11}\left( \frac{v+m}{2} + \frac{\tau}{4}; \frac{\tau}{2} \right) \nonumber \\
 &\times\sum_{\mu \in \mathbb{Z}} \exp\left( -\frac{\pi\tau_2}{2} \left( \mu + \frac{2\Im(u) - \Im(v)}{\tau_2} \right)^2 \right) \left( \mu - \frac{2\Im(u) - \Im(v)}{\tau_2} \right) (-1)^{\mu}\nonumber \\
 &\times \exp\left( -2\pi i (\mu + \frac{1}{2}) \left( u - \frac{v+m}{2} - \frac{\tau}{4} \right) \right) \exp\left( -\frac{\pi i \tau}{2} (\mu + \frac{1}{2})^2 \right) .
\end{align}
Now note that the factor of $\mu$ in the sum can arise from differentiating,
with respect to $u$, the exponential with an exponent linear in $\mu$. The structure of the sum
is also of the form of a theta function. After some manipulation we find
\begin{align}
\label{hol1b}
\frac{\partial}{\partial \overline{\tau}} \mathcal{R}_{1,1}(\tau, u, v)
&=\frac{\partial}{\partial u} \frac{1}{8\sqrt{2\tau_2}} e^{\pi i u} \exp\left( -\frac{\pi}{2\tau_2}\left( 2\Im(u) - \Im(v) - \frac{\tau_2}{2} \right)^2 \right) \nonumber \\
&\times \sum_{m=0}^1 
\vartheta_{11}\left( \frac{v+m}{2} + \frac{\tau}{4}; \frac{\tau}{2}
 \right) 
\vartheta_{11}\left( -\Re(u) + \frac{\Re(v)}{2} + \frac{\Re(\tau)}{4} + \frac{m}{2}; -\frac{\Re(\tau)}{2} \right).
\end{align}

Using some theta function identities, we can write the sum over $m$ of the
product of $\vartheta_{11}$-functions as products of $\vartheta_{00}$ and
$\vartheta_{01}$. The result is:
\begin{align}
\label{hol1c}
\frac{\partial}{\partial \overline{\tau}} \mathcal{R}_{1,1}(\tau, u, v)
 &=\frac{\partial}{\partial u} \frac{1}{8\sqrt{2\tau_2}} \exp\left( -\frac{\pi}{2\tau_2}\left( 2\Im(u) - \Im(v) \right)^2 \right) \nonumber \\
 &\times\sum_{m=0}^1 
\vartheta_{0m}\left( \frac{v}{2}; \frac{\tau}{2} \right) 
\vartheta_{0m}\left( -\Re(u) + \frac{\Re(v)}{2}; -\frac{\Re(\tau)}{2} \right).
\end{align}

Now we can simplify the notation a little by defining $z \equiv v - 2u$, and
using variables $z$ and $v$ we just replace $\frac{\partial}{\partial u}$ with
$-2 \frac{\partial}{\partial z}$. The result is
\begin{align}
\label{hol1d}
\frac{\partial}{\partial \overline{\tau}} \mathcal{R}_{1,1}(\tau, u, v)
 &=\frac{\partial}{\partial z} \frac{-1}{4\sqrt{2\tau_2}} \exp\left( -\frac{\pi}{2\tau_2}\left( \Im(z) \right)^2 \right)  \nonumber \\
 &\times\sum_{m=0}^1 \vartheta_{0m}\left( \frac{v}{2}; \frac{\tau}{2}
 \right) 
\vartheta_{0m}\left( \frac{\Re(z)}{2}; -\frac{\Re(\tau)}{2} \right). 
\end{align}

The most useful aspect of this notation is 
when we note that for $u = z_1$
and $v = z_1 + z_2 \equiv w$, 
and for $u = -z_2$ and $v = -z_1 - z_2 = -w$,
we have $z = z_2 - z_1$. 
So, in both cases we find (differing only in $v=w$ or
$v = -w$)
\begin{align}
\label{hol1e}
\left( \mathcal{D} + \frac{\Im(v) - 2 \Im(u)}{\tau_2} \right)
 \frac{\partial}{\partial \overline{\tau}} \mathcal{R}_{1,1}
&=\left( \frac{i}{\pi}\frac{\partial}{\partial z} + \frac{\Im(z)}{\tau_2} \right) \frac{\partial}{\partial z}
\frac{-1}{4\sqrt{2\tau_2}} \exp\left( -\frac{\pi}{2\tau_2}\left( \Im(z) \right)^2 \right) \nonumber \\
&\times\sum_{m=0}^1 
\vartheta_{0m}\left( \frac{\pm w}{2}; \frac{\tau}{2} \right) 
\vartheta_{0m}\left( \frac{\Re(z)}{2}; -\frac{\Re(\tau)}{2} \right)
\end{align}
but since $\vartheta_{0m}(-z; \tau) = \vartheta_{0m}(z; \tau)$ 
we get exactly the same expression in both cases. 
This means that when we calculate the
$\overline{\tau}$ derivative of the completed index (\ref{ModIndex}) the terms
arising from the $\overline{\tau}$ derivative of $\mathcal{R}_{1,1}$ in
(\ref{nonholA21}) cancel. So, we finally get the result which is indicative of
a recursion relation for the holomorphic anomaly:
\begin{align}
\label{holano1}
\frac{\partial}{\partial \overline{\tau}} \hat{\mathcal{I}}(\tau, z_1, z_2)
&=\frac{-i(z_2 - z_1)}{2 \tau_2^2}
 \frac{1}{\eta^4(\tau)}\frac{\vartheta_{11}^2(z_1; \tau)\vartheta_{11}^2(z_2;
 \tau)}
{\vartheta_{11}(z_1-z_2; \tau)\vartheta_{11}(z_1+z_2; \tau)}\nonumber \\
& \times\left( \hat{\mathcal{A}}_{1,1}(\tau, z_1, z_1+z_2) - \hat{\mathcal{A}}_{1,1}(\tau, -z_2, -z_1-z_2) \right). 
\end{align}

\subsection{Modular and elliptic transformations}
\label{subsectrans}
If we define
\begin{align}
\label{trans1a}
\Phi(\tau, z_1, z_2)&= \hat{\mathcal{A}}_{2,1}(\tau, z_1, z_1+z_2) - \hat{\mathcal{A}}_{2,1}(\tau, -z_2, -z_1-z_2), 
\end{align}
then we find the following transformation properties, noting that both
$\hat{\mathcal{A}}$ terms transform in the same way under these transformations:
\begin{align}
\label{trans1b}
\Phi(\tau + 1, z_1, z_2) &
=\Phi(\tau, z_1, z_2), &
\Phi(-\frac{1}{\tau}, \frac{z_1}{\tau}, \frac{z_2}{\tau})&
=\tau^2 e^{\frac{2\pi i}{\tau} z_1 z_2} \Phi(\tau, z_1, z_2), \\
\Phi(\tau, z_1+1, z_2) &=\Phi(\tau, z_1, z_2), &
\Phi(\tau, z_1+\tau, z_2) &=e^{-2\pi i z_2} \Phi(\tau, z_1, z_2), \\
\Phi(\tau, z_1, z_2+1) &=\Phi(\tau, z_1, z_2), &
\Phi(\tau, z_1, z_2+\tau)&=e^{-2\pi i z_1} \Phi(\tau, z_1, z_2). 
\end{align}
If we also include the theta and eta functions the index transforms as
\begin{align}
\label{trans1c}
\hat{\mathcal{I}}(\tau + 1, z_1, z_2)&
=e^{\frac{\pi i}{6}} \hat{\mathcal{I}}(\tau, z_1, z_2), &
\hat{\mathcal{I}}(-\frac{1}{\tau}, \frac{z_1}{\tau}, \frac{z_2}{\tau})&
=-i \tau e^{\frac{2\pi i}{\tau} z_1 z_2} \hat{\mathcal{I}}(\tau, z_1, z_2), \\
\hat{\mathcal{I}}(\tau, z_1+1, z_2)&
= \hat{\mathcal{I}}(\tau, z_1, z_2), &
\hat{\mathcal{I}}(\tau, z_1+\tau, z_2) &
=e^{-2\pi i z_2} \hat{\mathcal{I}}(\tau, z_1, z_2), \\
\hat{\mathcal{I}}(\tau, z_1, z_2+1) &= \hat{\mathcal{I}}(\tau, z_1, z_2),& 
\hat{\mathcal{I}}(\tau, z_1, z_2+\tau)&=e^{-2\pi i z_1} \hat{\mathcal{I}}(\tau, z_1, z_2). 
\end{align}

\subsection{Wall-crossing}
\label{subsubsecdec1}
The Appell-Lerch sum of order 2 is associated to meromorphic Jacobi forms of weight 2. 
It is shown in \cite{Dabholkar:2012nd} that any meromorphic Jacobi form $\varphi_{m}(\tau,z)$ with 
double poles at 
$z=z_{s}=\alpha\tau+\beta$, $\alpha,\beta\in S\subset\mathbb{Q}^{2}$ 
has a decomposition 
\begin{align}
\label{fourier1}
\varphi_{m}(\tau,z)&=\varphi_{m}^{F}(\tau,z)+\varphi_{m}^{P}(\tau,z). 
\end{align}
Here 
\begin{align}
\label{fourier2}
\varphi_{m}^{F}(\tau,z)&=
\sum_{l\in \mathbb{Z}/ 2m\mathbb{Z}}
h_{l}(\tau)\vartheta_{m,l}(\tau,z)
\end{align}
is a finite part and 
\begin{align}
\label{fourier3}
\varphi_{m}^{P}(\tau,z)&=
\sum_{s\in S/ \mathbb{Z}^{2}}
\left(
D_{s}(\tau)A_{1,m}^{s}(\tau,z)
+E_{s}(\tau)A_{2,m}^{s}(\tau,z)
\right)
\end{align}
is a polar part. Here $D_{s}(\tau)$ and $E_{s}(\tau)$ are residue functions defined by
\begin{align}
\label{fourier4}
e^{2\pi im\alpha z_{s}}
\varphi(\tau,z_{s}+\epsilon)
=\frac{E_{s}(\tau)}{(2\pi i\epsilon)^{2}}+\frac{D_{s}(\tau)-2m\alpha E_{s}(\tau)}{2\pi i\epsilon}+\mathcal{O}(1) ,
\end{align}
while $A_{1,m}^{s}(\tau,z)$ and $A_{2,m}^{s}(\tau,z)$ are universal Appell-Lerch sums \cite{Dabholkar:2012nd}
of order 1 and 2. 
In our analysis we saw multi-variable order 1 and 2 Appell-Lerch sums. 
In the single variable case these corresponding to taking $s=0$ above and are
defined by
\begin{align}
\label{apps1}
A_{1,m}(\tau,z)&=
-\frac12 \sum_{n\in \mathbb{Z}}q^{mn^{2}}x^{2mn}\frac{1+xq^{n}}{1-xq^{n}}, \\
\label{apps2}
A_{2,m}(\tau,z)&=
\sum_{n\in \mathbb{Z}}
\frac{q^{mn^{2}}x^{2mn+1}}{(1-xq^{n})^{2}}.
\end{align}

In the context of black hole microstate counting,  
the degeneracy of four-dimensional $\mathcal{N}=4$ quarter-BPS dyonic black holes with a
set of three fixed charges $(m,n,l)$ is given
by Fourier coefficients of the partition function, that is 
a meromorphic Siegel modular form of weight $-10$ 
\cite{Dijkgraaf:1996it, Shih:2005uc, David:2006yn}
\begin{align}
\label{bh1}
\mathcal{Z}_{\textrm{dyon}}&=\frac{1}{\Phi_{10}(\tau,z,\sigma)}
=\sum_{m=-1}^{\infty}\varphi_{m}(\tau,z)y^{m} .
\end{align}
Here $\Phi_{10}(\tau,z,\sigma)$ is the Igusa cusp form of weight $10$ 
and $\varphi_{m}(\tau,z)$ is a meromorphic Jacobi-form 
of weight $2$ and index $m$. 
According to the above decomposition theorem (\ref{fourier1}) of meromorphic Jacobi forms,  
$\varphi_{m}(\tau,z)$ in (\ref{bh1}) can be decomposed as
\begin{align}
\varphi_{m}(\tau,z)&=
\varphi_{m}^{F}(\tau,z)+\frac{p_{24}(m+1)}{\Delta(\tau)}A_{2,m}(\tau,z). 
\end{align}
Here the first term $\varphi_{m}^{F}(\tau,z)$ is a finite part without pole 
and counts the single-centered black holes 
while the second is a polar part with double poles 
and counts the multi-centered black holes that decay into its
single-centered constituents 
upon wall-crossing phenomena \cite{Dabholkar:2012nd}. 
In fact, the Appell-Lerch sum of order 2 is intimately related to an occurrence of 
wall-crossing due to its polar structure. 
To see this, it is useful to introduce an operation of averaging the residues at poles $z=z_{s}=\alpha+\beta\tau$
\begin{align}
\label{av1}
\mathrm{Av}^{(m)}\left[f(x)\right]
&:=\sum_{\lambda\in \mathbb{Z}}
q^{m\lambda^{2}}x^{2m\lambda} f(q^{\lambda}x).
\end{align}
This averaging operator constructs a Jacobi form of index $m$ 
out of an arbitrary function $f(x)$. 
Making use of the averaging operator, 
one can express the Appell-Lerch sum of order 2 as 
\begin{align}
\label{av2}
\mathrm{Av}^{(m)}\left[\frac{x}{(1-x)^{2}}\right]
&=A_{2,m}. 
\end{align}
The function $f(x)$ has an expansion
\begin{align}
\label{wc1}
\frac{x}{(1-x)^{2}}&=x+2x^{2}+3x^{3}+\cdots
\end{align}
in the range $|x|<1$ but it does not for $|x|>1$. 
This implies wall-crossing 
because different expansions of the meromorphic Jacobi form for 
$|x|<1$ and $|x|>1$ give different degeneracies as its coefficients. 
Correspondingly we have 
\begin{align}
\label{wc2}
A_{2,m}&=
\left(
\sum_{n\ge 0}{\sum_{l\ge0}}^{*}-\sum_{n<0}{\sum_{l\le0}}^{*}
\right)lq^{mn^{2}+ln}x^{2mn+l}
\end{align}
for $|q|<|x|<1$. 
Here ${\sum_{l} }^{*}$ is the sum for the term $l=0$ 
with multiplicity $\frac12$. 

These quarter-BPS black holes can be realized as 
a configuration of M2-M5 bound states in M-theory on 
$\textrm{K3}\times T^{2}$ \cite{Dabholkar:2012nd}.  
Let $T^{2}$ be a product of two circles 
$S^{1}_{\alpha}\times S^{1}_{\beta}$. 
Let $C^{1}$ be a homology 2-cycle of $T^{2}$ and 
$C^{2}$, $C^{3}$ be two 2-cycles in K3 which have intersection number 
\begin{align}
\label{ccc1}
\int_{\textrm{K3}\times T^{2}}C^{1}\wedge C^{2}\wedge C^{3}&=1. 
\end{align}
Let $\{D_{a}\}$ be 4-cycles dual to $\{C^{a}\}$, 
i.e. $D_{a}\cap C^{b}=\delta_{a}^{b}$. 
We consider the M2-M5 bound states with 
$w$ units of momentum along M-circle $S_{0}^{1}$ 
where $\widetilde{K}$ units of M5-brane charge wrap $D_{1}\times S_{0}^{1}$, 
$Q_{1}$ units of M5-brane charge wrap $D_{2}\times S^{1}_{0}$, 
$Q_{5}$ units of M5-brane charge wrap $D_{3}\times S^{1}_{0}$ and 
$\widetilde{n}$ units of M2-brane charge wrap $T^{2}$. 
Then 
\begin{align}
\label{mnl1}
m&=Q_{1}Q_{5},& 
n&=w\widetilde{K},& 
l&=\widetilde{n}\widetilde{K}
\end{align}
can be identified with the charges 
of the quarter-BPS dyonic black hole states,
and the number of BPS bound states of the brane configuration can be viewed 
as the degeneracy of the black holes. 
When the M5-brane charges $\widetilde{K}, Q_{1}$ and $Q_{5}$ are fixed, 
the charges $(m,n,l)$ of the black hole 
are determined by the momentum $w$ and the M2-brane charge $\widetilde{n}$ 
which would be specified by the quantum numbers
of the derivation $d=-L_{0}$ and those of Cartan elements of the
Lie superalgebra $\mathfrak{sg}$ respectively. 
Hence, under certain circumstances
our index would have an interpretation in terms of black hole microstate counting.  
An appearance of the second order multi-variable Appell-Lerch sum $\mathcal{A}_{2,m}$ 
would suggest that such multi-centered black holes may decay into single-centered black holes 
\cite{Denef:2000nb, Denef:2002ru, Bates:2003vx, Denef:2007vg}.

From the perspective of the M2-M5 system it is expected that 
wall-crossing occurs due to the configuration of stretched M2-branes so that 
the moduli space of the M2-M5 system may develop a new branch 
at a particular critical value of the $C$-field on the M5-branes.

\section{Discussion}
%
\label{secdis}
We have described BPS indices for supergroup WZW models which we have argued
count the degeneracies of BPS states of the intersecting M2-M5 system considered in \cite{Okazaki:2015fiq}. 
The BPS states are specified by the highest weight modules of the affine Lie superalgebra 
in such a way that 
the number of stretched M2-branes is equal to the degree of atypicality. In addition, 
the momenta along a wrapped circle are given by the Virasoro modes that amount to the derivation,
and the M2-brane charges under the $C$-fields are given by the Cartan elements of the finite Lie superalgebra. 
When all these M2-branes are sandwiched between the M5-branes, 
in which case the BPS states are the modules with maximal atypicality, 
the indices can be evaluated using the Kac-Wakimoto character formula 
\cite{MR1810948, MR3200431}. 
Quite remarkably they are written in terms of the $q$-series known as 
Ramanujan's mock theta functions 
\cite{MR2280843, MR1567721, MR1542896}. 
Our result is an encounter of the mock Jacobi forms in the BPS indices of the M-strings, 
which are defined in the supergroup WZW models in the same manner 
as the equivariant elliptic genus studied in 
\cite{Haghighat:2013gba, Haghighat:2013tka, Hohenegger:2013ala, Haghighat:2014pva, Hohenegger:2015cba}. 
The indices have a structure which suggests there is wall-crossing 
in the BPS state counting of the M2-M5 system, related to
universal features of the Appell-Lerch sums.
We have argued that 
the mock modularity of the supercharacters of affine Lie superalgebras
reflects the non-holomorphic atypical sector of the Hilbert space of the
supergroup WZW models. 
To obtain the non-holomorphic modular parts of 
the torus partition function of supergroup WZW models, 
we have invoked Zwegers' method \cite{Zwegers:2008zna}, closely following the
discussion in \cite{Dabholkar:2012nd} and particularly \cite{Ashok:2014nua}.

There are many future directions to consider.
%
Clearly it is desirable to extend our explicit evaluation of the indices for 
$PSL(2|2)_{k=1}$ to other cases. 
The indices reduce to a specialization of 
the supercharacters of integrable highest weight modules 
over affine Lie superalgebras. However, at present 
explicit calculation of supercharacters is only available 
for $\widehat{\mathfrak{gl}}(N|1)$ and 
$\widehat{\mathfrak{sl}}(N|1)$ in \cite{MR1810948}, 
for $\widehat{\mathfrak{psl}}(2|2)$ in \cite{MR3200431}, 
for $\widehat{\mathfrak{osp}}(3|2)$ in \cite{MR3534829}, and
for some general basic Lie algebras $\widehat{\mathfrak{sg}}$ 
in \cite{MR3535359}. 
The case of most relevance for our application is $\widehat{\mathfrak{gl}}(N|N)$
which arises in the case of $N$ M2-branes between the M5-branes. Understanding
the dependence of the spectrum on $N$ is an obvious issue, and perhaps some
aspects can be studied even without the complete explicit expression for the
supercharacter.

%
Going beyond the M-brane configurations considered in \cite{Okazaki:2015fiq}
we could consider configurations with M2-branes on both sides of an M5-brane
and more than one M5- and M5$'$-brane. In the case of parallel M5-branes the
index has been calculated \cite{Haghighat:2013gba}
using various techniques including topological strings.
The type IIB description of such systems in flat space has been considered
by Niarchos \cite{Niarchos:2015lla} by the addition of D5-branes to the
ABJM configuration. We have commented on the description with both M5- and
M5$'$-branes, including either D5- and D5$'$-branes or NS5- and NS5$'$-branes
in type IIB. We expect this will lead to further understanding of the
M2-M5 system, with or without the topological twisting. Certainly, as we
discussed, we expect this to lead to an understanding of the detailed
coupling between ABJM models describing M2-branes on either side of an M5-brane.
In the type IIB configuration this can be studied in terms of open strings
connecting the D3-branes 
and recent works \cite{Bullimore:2016nji, Chung:2016pgt} on 
the supersymmetric boundary conditions in three-dimensional $\mathcal{N}=4$ gauge theories 
will play a key role to give the description of these brane tiling models 
as two-dimensional gauge theories. 
In the case of supergroup WZW models, 
we expect that this would give a specific model based on $GL(N|N) \times GL(M|M)$. It may also be possible
to extend this analysis in type IIB to include generalizations of the ABJM
model, such as those based on the ABJ theory or with orthogonal and symplectic
gauge groups \cite{Aharony:2008gk, Hosomichi:2008jb, deMedeiros:2008zh}. 
Although it is not clear how to relate all these cases to
M-brane configurations, we would expect some (but not all) to correspond to supergroup WZW models.

The Appell-Lerch sums, which we have found in the indices, are known to play 
an important role in mathematics and physics.
In particular, they appear as the Fourier coefficients of the generating functions 
in various counting problems. 
We expect that the appearance of these sums from M-brane constructions will
lead to a more unified formalism,
relating different aspects of the Appell-Lerch sums. 
To seek gauge theoretical descriptions, 
we could start from the world-volume theory of M5-branes 
wrapping a 4-manifold to obtain four-dimensional twisted $\mathcal{N}=4$ gauge theories 
\cite{Minahan:1998vr,Alim:2010cf}. 
In \cite{Manschot:2014cca} 
the generating function of topological invariants 
of the moduli space of vector bundles over 4-manifolds was
evaluated as the partition function of four-dimensional twisted $\mathcal{N}=4$ 
gauge theories, which is expressed in terms of
multi-variable Appell-Lerch sums.
Also, in the weak string coupling region, 
one could calculate indices in the world-volume theory of branes 
as the generating functions of certain topological invariants. 
In \cite{MR3433756} the generating functions 
of Gromov-Witten invariants of elliptic orbifolds are given 
by multi-variable Appell-Lerch sums. 
In the strong string coupling region, 
the brane system would involve the gravitational interaction 
and the indices would count the microstates of the black holes. 
As we have seen, the partition functions of the multi-centered black holes are expressed 
in terms of the Appell-Lerch sums \cite{Dabholkar:2012nd}.  
We hope to report on progress from these view points 
in subsequent works. 
%

\subsection*{Acknowledgements}
We would like to thank 
Thomas Creutzig, Kazuo Hosomichi and Katsushi Ito 
for useful discussions and comments. 
We also thank Minoru Wakimoto and Shun-Jen Cheng for helpful explanation about representation theory of Lie superalgebra 
and sharing valuable mathematical ideas. 
TO thanks the organizers of the 
\textit{Quantum Geometry, Duality and Matrix Models} in Moscow 
and 
the organizers of the 
\textit{NCTS Annual Theory Meeting 2016: Particles, Cosmology and
String} in Hsinchu 
for the opportunity to talk about the results of the present paper.
TO is supported by MOST under the Grant No.105-2811-066. 
DJS is supported in part by the STFC Consolidated Grant ST/L000407/1.

\appendix

\section{Modular forms}
\label{appjacobi}
The Dedekind eta function
\begin{align}
\label{eta1a}
\eta(\tau)&=q^{\frac{1}{24}}
\prod_{n=1}^{\infty}(1-q^{n})
\end{align}
satisfies the modular transformation properties
\begin{align}
\label{eta1b}
\eta\left(
-\frac{1}{\tau}
\right)&=(-i\tau)^{\frac12} \eta(\tau),& 
\eta(\tau+1)&
=e^{\frac{\pi i}{12}}\eta(\tau).
\end{align}
The four Jacobi theta functions are defined by \cite{MR2352717}
\begin{align}
\label{theta1a}
\vartheta_{00}(\tau,z)
&=\vartheta(\tau,z)=\sum_{n\in \mathbb{Z}} e^{2\pi inz}q^{\frac{n^{2}}{2}}
=\prod_{n=1}^{\infty}
(1-q^{n})(1+e^{2\pi iz}q^{n-\frac12})(1+e^{-2\pi iz}q^{n-\frac12}),\\
\label{theta1b}
\vartheta_{01}(\tau,z)
&=\vartheta_{00}(\tau,z+\frac12)
=\prod_{n=1}^{\infty}
(1-q^{n})(1-e^{2\pi iz}q^{n-\frac12})(1-e^{-2\pi iz}q^{n-\frac12}),\\
\label{theta1c}
\vartheta_{10}(\tau,z)
&=e^{\frac{\pi i\tau}{4}}e^{\pi iz}\vartheta_{00}(\tau,z+\frac{\tau}{2})
=e^{\frac{\pi i\tau}{4}}e^{-\pi iz}
\prod_{n=1}^{\infty}
(1-q^{n})(1+e^{2\pi iz}q^{n-1})(1+e^{-2\pi iz}q^{n}),\\
\label{theta1d}
\vartheta_{11}(\tau,z)
&=ie^{\frac{\pi i\tau}{4}}e^{\pi iz}\vartheta_{00}(\tau,z+\frac{\tau}{2}+\frac12)
=e^{\frac{\pi i\tau}{4}}e^{-\pi i(z+\frac12)}
\prod_{n=1}^{\infty}
(1-q^{n})(1-e^{2\pi iz}q^{n-1})(1-e^{-2\pi iz}q^{n}).
\end{align}
We have the transformation laws 
\begin{align}
\label{theta2a}
\vartheta_{00}\left(-\frac{1}{\tau},\frac{z}{\tau}\right)&=
(-i\tau)^{\frac12}e^{\frac{\pi iz^{2}}{\tau}}\vartheta_{00}(\tau,z),&
\vartheta_{01}\left(-\frac{1}{\tau},\frac{z}{\tau}\right)&=
(-i\tau)^{\frac12}e^{\frac{\pi iz^{2}}{\tau}}\vartheta_{10}(\tau,z),\\
\label{theta2c}
\vartheta_{10}\left(-\frac{1}{\tau},\frac{z}{\tau}\right)&=
(-i\tau)^{\frac12}e^{\frac{\pi iz^{2}}{\tau}}\vartheta_{01}(\tau,z),& 
\vartheta_{11}\left(-\frac{1}{\tau},\frac{z}{\tau}\right)&=
(i\tau)^{\frac12}e^{\frac{\pi iz^{2}}{\tau}}\vartheta_{11}(\tau,z),
\end{align}
and 
\begin{align}
\label{theta3a}
\vartheta_{00}(\tau+1,z)&=\vartheta_{01}(\tau,z),& 
\vartheta_{01}(\tau+1,z)&=\vartheta_{00}(\tau,z),\\
\label{theta3c}
\vartheta_{10}(\tau+1,z)&=e^{\frac{\pi i}{4}}\vartheta_{10}(\tau,z),&
\vartheta_{11}(\tau+1,z)&=e^{\frac{\pi i}{4}}\vartheta_{11}(\tau,z).
\end{align}


\bibliographystyle{utphys}
\bibliography{ref}

\end{document}